\documentclass[prd,aps,floats,amssymb,preprint,nofootinbib]{revtex4}

\usepackage{epsfig}
\usepackage{amssymb}
\usepackage{bbm}
\usepackage{color}
\usepackage{ulem}

\usepackage{enumitem}
\usepackage{graphicx}
\usepackage{subfigure}
\usepackage{amsmath}
\usepackage{bbm}
\usepackage{color}

\def\be{\begin{equation}}
\def\ee{\end{equation}}
\def\bea{\begin{eqnarray}}
\def\eea{\end{eqnarray}}

\providecommand{\lsim}{\lesssim}

\begin{document}


\title{Screening bulk curvature in the presence of large brane tension}
\author{Nishant Agarwal$^1$}
\author{Rachel Bean$^1$}
\author{Justin Khoury$^2$}
\author{Mark Trodden$^2$}

\affiliation{$^1$Department of Astronomy, Cornell University, Ithaca, New York 14853, USA \\
$^2$Center for Particle Cosmology, Department of Physics and Astronomy, University of Pennsylvania,
Philadelphia, Pennsylvania 19104, USA}

\date{\today}

\begin{abstract}
We study a flat brane solution in an effective $5D$ action for cascading gravity and propose a mechanism to screen extrinsic curvature in the presence of a large tension on the brane. The screening mechanism leaves the bulk Riemann-flat, thus making it simpler to generalize large extra dimension dark energy models to higher codimensions. By studying an action with cubic interactions for the brane-bending scalar mode, we find that the perturbed action suffers from ghostlike instabilities for positive tension, whereas it can be made ghost-free for sufficiently small negative tension.
\end{abstract}

\maketitle


\section{Introduction}
\label{intro}

The problem of cosmic acceleration and its possible explanation as a cosmological constant have led to a wide variety of models in theoretical physics. Higher-dimensional theories of dark energy, in which our Universe is viewed as a $4D$ brane living in a higher-dimensional bulk, offer an interesting proposal towards understanding dark energy as a manifestation of the presence of extra dimensions of space-time. Much progress has been made in this field using the braneworld picture in which all standard model particles are confined to a $4D$ brane, while gravity is free to explore the bulk \cite{Lukas:1998yy,ArkaniHamed:1998rs,Antoniadis:1998ig}. This makes it possible to have cosmologically large extra dimensions \cite{ArkaniHamed:1998rs,Randall:1999ee,Randall:1999vf}. The Dvali-Gabadadze-Porrati (DGP) model \cite{Dvali:2000hr}, in particular, takes this idea to the extreme and considers our $4D$ Universe to be embedded in a $5D$ bulk of {\it infinite} extent. Despite being observationally disfavored \cite{Rydbeck:2007gy,Fang:2008kc,Lombriser:2009xg,Guo:2006ce}, the normal branch of the DGP model is perturbatively ghost-free, in contrast to the self-accelerating branch \cite{Nicolis:2004qq,Koyama:2005tx,Gorbunov:2005zk,Charmousis:2006pn,Dvali:2006if,Gregory:2007xy}, and thus represents a perturbatively consistent infrared modification of gravity in which the graviton has a soft mass.

Infinitely large extra dimensions also offer a promising arena for realizing Rubakov and Shaposhnikov's proposal~\cite{Rubakov:1983bz} for addressing the cosmological constant problem, namely that brane tension could curve the extra dimensions while leaving the $4D$ geometry flat. While tantalizing, this idea immediately fails if the extra dimensions are compactified, since $4D$ general relativity, and hence standard no-go arguments \cite{Weinberg:1988cp}, apply below the compactification scale. Moreover, obtaining a flat $4D$ geometry with compact extra dimensions requires canceling the brane tension against other branes and/or bulk fluxes \cite{Gibbons:2000tf}. The situation is more promising if the extra dimensions have infinite volume. The weakening of gravity as it enters the higher-dimensional regime (combined with an intrinsic curvature term on the brane) at least suggests that vacuum energy, by virtue of being the longest-wavelength source, might only appear small because it is {\it degravitated} \cite{Dvali:2002pe,ArkaniHamed:2002fu,Dvali:2007kt}.

The generalization of large extra-dimension dark energy models to higher codimensions is important not only for the cosmological constant problem but also for their possible embedding into string theory \cite{Dvali:2002pe,Dvali:2000xg}. Previous attempts of such a generalization have been found to give rise to a divergent brane-to-brane propagator and ghost instabilities around flat space \cite{Dubovsky:2002jm,Gabadadze:2003ck}. Furthermore, for a static bulk, the geometry for codimension $N > 2$ has a naked singularity at a finite distance from the brane, for arbitrarily small tension \cite{Dvali:2002pe}.

The cascading gravity framework~\cite{deRham:2007xp,deRham:2007rw,Corradini:2007cz,Corradini:2008tu,deRham:2009wb,deRham:2010rw} avoids these pathologies by embedding the $4D$ brane within a succession of higher-dimensional branes, each with their own intrinsic curvature term. The brane-to-brane propagator is regulated by the intrinsic curvature term of the higher-dimensional brane. Meanwhile, in the simplest codimension-2 case, consisting of a $4D$ brane embedded in a $5D$ brane within a $6D$ bulk, the ghost is cured by including a sufficiently large tension $\Lambda$ on the (flat) $4D$ brane:
\be
	\Lambda \geq \frac{2}{3}m_6^2M_4^2\,,
\label{ghostfreebound}
\ee
where $m_6 \equiv M_6^4/M_5^3$, and $M_D$ denotes the Planck mass in $D$ dimensions. This stability bound was first derived through the decoupling limit $M_5,M_6\rightarrow \infty$, keeping the strong-coupling scale $\Lambda_6 = (m_6^4M_5^3)^{1/7}$ fixed. In this limit, the $6D$ framework reduces to a local theory on the $5D$ brane, describing weak-field $5D$ gravity coupled to a self-interacting scalar field $\pi$. The bound~(\ref{ghostfreebound}) was confirmed in \cite{deRham:2010rw} through a complete perturbation analysis in the full $6D$ set-up.

The codimension-2 solution exhibits degravitation: the brane tension creates a deficit angle in the bulk, leaving the geometry flat. Since the deficit angle must be less than $2\pi$, the tension is bounded from above:
\be
	\Lambda \leq 2\pi M_6^4\,.
\label{geometricalbound}
\ee
Since $M_6$ is constrained phenomenologically to be less than $\sim$meV, this upper bound is unfortunately comparable to the dark energy scale. 
Given its geometrical nature, however, this is likely an artifact of the codimension-2 case and is expected to be absent in higher codimensions. This motivated \cite{deRham:2009wb} to study the codimension-3 case, consisting of a $4D$ brane living on a $5D$ brane, itself embedded in a $6D$ brane, together in a $7D$ bulk space-time. In the limit of small tension on the $4D$ brane, such that the weak-field approximation is valid, \cite{deRham:2009wb} showed that the bulk geometry is {\it non-singular} everywhere (away from the brane) and asymptotically flat, with the induced $4D$ geometry also flat.

In a recent paper \cite{Agarwal:2009gy}, we proposed a proxy theory for the full $6D$ cascading gravity model by covariantizing the $5D$ effective theory obtained through the decoupling limit. The resulting action is a $5D$ scalar-tensor theory, describing $5D$ gravity and the brane-bending scalar mode (denoted by $\pi$), coupled to a $4D$ brane. The scalar field is of the conformal galileon type \cite{Nicolis:2008in}, with a cubic self-interaction term \cite{Luty:2003vm,Nicolis:2004qq}. Since our brane is a codimension-1 object in this case, the equations of motion are more tractable and allowed us in \cite{Agarwal:2009gy} to derive a rich cosmology on the brane. A similar strategy was used in earlier work \cite{Chow:2009fm} to construct an effective $4D$ covariant theory, which was shown to faithfully reproduce much of the phenomenology of the full $5D$ DGP model. See \cite{Deffayet:2009wt,Silva:2009km,DeFelice:2010pv,Mota:2010bs} for related work.

The goal of this paper is to explore whether this effective framework also allows for flat brane solutions with tension and, if so, whether such degravitated solutions are stable. In particular, are the bounds (\ref{ghostfreebound}) and (\ref{geometricalbound}) reproduced in the effective theory?

Remarkably, we find that our $5D$ theory allows for flat brane solutions for {\it arbitrarily large} tension, with the bulk geometry being non-singular. The cascading origin of the theory is essential to the viability of these solutions: if we let $m_6 \rightarrow \infty$, corresponding to turning off the cubic scalar self-interaction, the bulk geometry develops a naked singularity a finite distance from the brane, as in \cite{Kachru:2000hf}.

Our mechanism for screening the brane cosmological constant relies crucially on $\pi$. In order for the theory to have a well-defined variational principle, the cubic self-interaction term requires appropriate interactions for $\pi$ on the brane, analogous to the Gibbons-Hawking-York term for gravity. In the presence of brane tension, these scalar boundary terms screen the tension, resulting in a flat geometry. This is the interpretation of our mechanism in the Jordan frame, in which the scalar is non-minimally coupled to gravity. There is of course a similar intuitive explanation in the 
Einstein frame. There, based on the Israel junction conditions, one would expect that a large brane tension should imply large extrinsic curvature, and hence large ({\it i.e.} super-Planckian) bulk curvature near the brane. Instead, the scalar boundary terms effectively screen the tension, much like the screening of charges in a dielectric medium, resulting in a small source for bulk gravity. 

The screening mechanism we propose seems to resolve the problem with earlier self-tuning attempts. A perturbative analysis of this mechanism, however, shows that it is difficult to avoid ghosts in such a model for positive brane tension, while it is possible to obtain consistent ghost-free solutions for negative tension. We further find that the model is free of gradient instabilities, and scalar perturbations propagate sub-luminally along the extra dimension. It is also worth mentioning that we only consider solutions in which the bulk is flat, hence we are working on a different branch of solutions than those studied in \cite{Agarwal:2009gy}, and our results are in no way contradictory to \cite{deRham:2010rw,Agarwal:2009gy}.

We have organized our paper in the following way. After briefly reviewing cascading gravity in Sec.~\ref{cascading}, we present the flat brane solution in Sec.~\ref{flatbrane}. In Sec.~\ref{ghosts} we discuss perturbations to the screening solution around a flat background, and derive various conditions for stability, both in the bulk and on the brane. We summarize our results and discuss future research avenues in Sec.~\ref{conclusions}.

A comment on our notation: We use the mostly positive signature convention. Indices $M,N,...$ run over $0,1,2,3,5$ ({\it i.e.} the $4+1D$ coordinates) and indices $\mu,\nu,...$ run over $0,1,2,3$ ({\it i.e.} the $3+1D$ coordinates). We denote the fifth dimensional coordinate by $y = x^{5}$.


\section{Overview of cascading gravity}
\label{cascading}

Consider a $6D$ cascading gravity model in which a 3-brane is embedded in a succession of higher-dimensional 
branes, each with its own Einstein-Hilbert action~\cite{deRham:2007xp,deRham:2007rw},
\bea
\nonumber
	S_{\rm cascade} & = & \int_{\rm bulk} {\rm d}^{6}x\sqrt{-g_{6}}\frac{M_{6}^{4}}{2}R_{6} + \int_{\rm 4-brane} {\rm d}^{5}x \sqrt{-g_{5}} \frac{M_{5}^{3}}{2}R_{5} \nonumber \\
	& & + \int_{\rm 3-brane} {\rm d}^{4}x \sqrt{-g_{4}} \left(\frac{M_{4}^{2}}{2}R_{4} + {\cal L}_{\rm matter}\right)\,, \ \ \ 
\label{S6}
\eea
where, as mentioned earlier, $M_D$ denotes the Planck mass in $D$ dimensions.
The gravitational force law on the 3-brane ``cascades" from $1/r^{2}$ to $1/r^{3}$ and from $1/r^{3}$ to $1/r^{4}$ as the Universe transitions from $4D$ to $5D$ and ultimately to $6D$ at the crossover scales $m_{5}^{-1}$ and $m_{6}^{-1}$ respectively, where\footnote{Strictly speaking, the $4D\rightarrow 5D\rightarrow 6D$ cascading behavior of the force law requires $m_5^{-1} < m_6^{-1}$, thereby allowing for an intermediate $5D$ regime. If $m_5^{-1} > m_6^{-1}$, on the other hand, the scaling of the force law transitions directly from $1/r^2$ to $1/r^4$ at the crossover scale $m_6^{-1}$.}
\bea
	m_{5} = \frac{M_{5}^{3}}{M_{4}^{2}}\,, \ \ \ \ \ m_{6} = \frac{M_{6}^{4}}{M_{5}^{3}} \ .
\label{m5m6}
\eea
As mentioned in Sec.~\ref{intro}, this theory allows for degravitated solutions ---  a 3-brane with tension creates a deficit angle in the bulk
while remaining flat. Furthermore, the theory is perturbatively ghost-free provided the 3-brane tension is sufficiently large that~(\ref{ghostfreebound})
is satisfied.

In the decoupling limit $M_{5}, \ M_{6} \rightarrow \infty$, with the strong-coupling scale
\be
	\Lambda_{6} = (m_{6}^{4}M_{5}^{3})^{1/7}
\label{lambda6}
\ee
held fixed, we can expand the action (\ref{S6}) around flat space and integrate out the sixth dimension \cite{Luty:2003vm,Agarwal:2009gy}. The resulting action is local in $5D$ and describes 
weak-field gravity coupled to a scalar degree of freedom $\pi$:
\begin{eqnarray}
	S_{\rm decouple} & = & \frac{M_{5}^{3}}{2} \int_{\rm bulk} {\rm d}^{5}x \left[ -\frac{1}{2} h^{MN}(\mathcal{E}h)_{MN} + \pi\eta^{MN}(\mathcal{E}h)_{MN} - \frac{27}{16m_6^2} (\partial\pi)^{2} \Box_{5}\pi \right] \nonumber \\
	& & + \int_{\rm brane} {\rm d}^{4}x \left[ -\frac{M_{4}^{2}}{4} h^{\mu\nu}(\mathcal{E}h)_{\mu\nu} + \frac{1}{2} h^{\mu\nu}T_{\mu\nu} \right]\,,
\label{5ddec}
\end{eqnarray}
where $(\mathcal{E}h)_{MN} =  -\Box_{5} h_{MN}/2 + \ldots$ is the linearized Einstein tensor. The scalar $\pi$ is the helicity-0 mode of the massive spin-2 graviton on the 4-brane and measures the extrinsic curvature of the 4-brane in the $6D$ bulk space-time. An obvious advantage offered by the decoupling theory is that the 3-brane now represents a codimension-1 object, which greatly simplifies the analysis. On the other hand, its regime of validity is of course restrained to the weak-field limit and therefore of limited interest for obtaining cosmological or degravitated solutions.

In~\cite{Agarwal:2009gy}, we proposed a proxy theory for the full $6D$ cascading gravity model by extending~(\ref{5ddec}) to a fully covariant,
non-linear theory of gravity in $5D$ coupled to a 3-brane,
\bea
	S & = & \frac{M_{5}^{3}}{2}\int_{\rm bulk}{{\rm d}^{5}x\sqrt{-g_{5}}\left[\Omega(\pi) R_{5} - \frac{27}{16m_6^2}(\partial\pi)^2 \Box_{5}\pi \right]} \nonumber \\
	& & + \int_{\rm brane}{{\rm d}^{4}x\sqrt{-g_{4}}\left( \frac{M_{4}^{2}}{2}R_{4} + \mathcal{L}_{\rm{matter}} \right)}.
\label{5dcov}
\eea
This reduces to~(\ref{5ddec}) in the weak-field limit provided that $\Omega(\pi) \approx 1 - 3\pi/2$ for small $\pi$. In~\cite{Agarwal:2009gy}, we chose
$\Omega(\pi) = e^{-3\pi/2}$ and derived the induced cosmology on a moving 3-brane in static bulk space-time solutions. Interestingly, this choice
corresponds in Einstein frame to the $5D$ generalization of the cubic conformal galileon~\cite{Nicolis:2008in}, whose structure is protected by symmetries. While the proposed covariantization of~(\ref{5ddec}) is by no means unique, our hope is that~(\ref{5dcov}) captures the salient features of the $6D$ cascading gravity model, and furthermore that the resulting predictions are at least qualitatively robust to generalizations of~(\ref{5dcov}). 

In this paper, we want to address whether~(\ref{5dcov}) allows the 3-brane to have tension while remaining flat. To parallel the corresponding $6D$ solutions, where the bulk acquires a deficit angle while remaining flat, we will impose that the $5D$ (Jordan-frame) metric is Minkowski space. For most of the analysis, we will leave $\Omega(\pi)$ as a general function, and derive constraints on its form based on stability requirements.

We work in the ``half-picture'', in which the brane is a boundary of the bulk space-time. In this case, the action~(\ref{5dcov}) is not complete without the appropriate Gibbons-Hawking-York (GHY) terms on the brane \cite{York:1972sj,Gibbons:1976ue}, both for the metric and for $\pi$~\cite{Dyer:2009yg}, to ensure a well-defined variational principle. These were derived in flat space in~\cite{Dyer:2009yg} and around a general backgroud in~\cite{Agarwal:2009gy}, and the complete $5D$ action is
\begin{eqnarray}
	S & = & \frac{M_{5}^{3}}{2} \int_{\rm bulk} {\rm d}^{5}x \sqrt{-g_{5}}\Omega \left( R_{4} + K^{2} - K_{\mu\nu}K^{\mu\nu} + 2K\frac{{\mathcal L}_{n}\Omega}{\Omega} - 2\frac{\Box_{4}\Omega}{\Omega} \right) \nonumber \\
	& & - \ \frac{27M_{5}^{3}}{32m_{6}^{2}} \int_{\rm bulk} {\rm d}^{5}x \sqrt{-g_{5}} (\partial\pi)^{2} \Box_{5}\pi - \frac{27M_{5}^{3}}{32m_{6}^{2}} \int_{\rm brane} {\rm d}^{4}x \sqrt{-q} \left( \partial_{\mu}\pi \partial^{\mu}\pi {\mathcal L}_{n}\pi + \frac{1}{3}({\mathcal L}_{n}\pi)^{3} \right) \nonumber \\
	& & + \ \frac{1}{2}\int_{\rm brane}{{\rm d}^{4}x\sqrt{-q}\left( \frac{M_{4}^{2}}{2}R_{4} + \mathcal{L}_{\rm{matter}} \right)} \,.
\label{5dcovcomplete}
\end{eqnarray}
Here $q_{\mu\nu} = g_{\mu\nu} - n_\mu n_\nu$ is the $4D$ induced metric, and $K_{\mu\nu} \equiv {\cal L}_{n} q_{\mu\nu}/2$ is the extrinsic curvature of the brane, where $n^\alpha$ is the unit normal to the brane, and ${\cal L}_{n}$ is the Lie derivative with respect to the normal. Note that we have added an extra factor of $1/2$ in the brane action so that the Israel junction conditions obtained using (\ref{5dcovcomplete}) match with those obtained in the ``full-picture''. The assumed $\mathbb{Z}_{2}$ symmetry across the brane guarantees that the bulk action in $y \geq 0$ is equal to that in $y \leq 0$, while the bulk in (\ref{5dcovcomplete}) is defined only in $y \geq 0$.

Varying (\ref{5dcovcomplete}) with respect to the metric leads to the Einstein field equations,
\begin{eqnarray}
	\Omega G_{MN} & = & -\frac{27}{16m_6^2} \Bigg[ \partial_{(M}(\partial\pi)^{2}\partial_{N)}\pi - \ \frac{1}{2}g_{MN}\partial_{K}(\partial\pi)^{2}\partial^{K}\pi - \partial_{M}\pi\partial_{N}\pi\Box_{5}\pi \Bigg] \nonumber \\
	& & - \left( g_{MN}\Box_{5} - \nabla_{M}\nabla_{N} \right) \Omega \ ,
\label{EFeqn}
\end{eqnarray}
where $G_{MN}$ is the $5D$ Einstein tensor, and parentheses around indices denote symmetrization: $X_{(MN)} \equiv (X_{MN} + X_{NM})/2$. The matter stress-energy tensor on the brane is defined as
\be
	T_{\mu\nu}^{(4)} \equiv - \frac{2}{\sqrt{-q}} \frac{\delta (\sqrt{-q}{\cal L}_{\rm matter})}{\delta q^{\mu\nu}} \ .
\ee
Similarly, varying with respect to $\pi$ gives us the $\pi$ equation of motion,
\begin{equation}
	(\Box_{5}\pi)^{2} - (\nabla_{M}\partial_{N}\pi)^{2} - R^{MN}_5 \partial_{M}\pi \partial_{N}\pi  = -\frac{8}{27}m_{6}^{2}\Omega_{,\pi}R_{5} \ ,
\label{pieom}
\end{equation}
with $\Omega_{,\pi} \equiv {\rm d}\Omega/{\rm d}\pi$. We further obtain the Israel junction conditions at the brane position by setting the boundary contributions to the variation of the action (\ref{5dcovcomplete}) to zero. Variation with respect to the metric gives us the Israel junction condition
\begin{eqnarray}
	2M_{5}^{3} \Omega \left( K q_{\mu\nu} - K_{\mu\nu} + \frac{\Omega_{,\pi}}{\Omega} q_{\mu\nu} {\cal L}_{n} \pi \right) & = & \frac{27M_{5}^{3}}{8m_{6}^{2}} \left( \partial_{\mu}\pi \partial_{\nu}\pi {\cal L}_{n} \pi + \frac{1}{3} q_{\mu\nu} \left( {\cal L}_{n}\pi \right)^{3} \right) \nonumber \\
	& & + \ T^{(4)}_{\mu\nu} - M_4^2G_{\mu\nu}^{(4)} \ ,
	\label{covjc1}
\end{eqnarray}
while varying with respect to $\pi$ yields the scalar field junction condition
\be
	\Omega_{,\pi}K - \frac{27}{16m_{6}^{2}} \Big( K_{\mu\nu} \partial^{\mu}\pi \partial^{\nu}\pi + 2{\cal L}_{n}\pi \Box_{4}\pi + K ({\cal L}_{n} \pi)^{2} \Big) = 0 \ .
\label{covjc2}
\ee

In the balance of this paper we seek flat brane solutions to the bulk equations~(\ref{EFeqn}) and~(\ref{pieom}), with boundary conditions set by~(\ref{covjc1}) and~(\ref{covjc2}).


\section{Obtaining flat brane solutions for any tension}
\label{flatbrane}

In this section we seek flat 3-brane solutions to the above equations of motion. To mimic the $6D$ situation where the brane remains flat but creates a deficit angle in a flat $6D$ bulk, we impose that the $5D$ (Jordan-frame) geometry is Minkowski space:
\be
{\rm d}s^{2}_{\rm bulk} = \eta_{MN}{\rm d}x^M {\rm d}x^N = -{\rm d}\tau^2 + {\rm d}\vec{x}^2 + {\rm d}y^2 \,.
\label{flatbulk}
\ee
Similarly, the induced metric on the brane should also be flat. By Lorentz invariance, clearly we can assume the brane to be at fixed position, $y=0$, with the extra dimension therefore extending from $y=0$ to $\infty$. By symmetry, we also have $\pi = \pi(y)$.

With these assumptions, the $(5,5)$ component of the field equations (\ref{EFeqn}) and the $\pi$ equation of motion (\ref{pieom}) are trivially satisfied, while the $(\mu,\nu)$ components of (\ref{EFeqn}) reduce to
\begin{eqnarray}
	\pi'' & = & \frac{\Omega_{,\pi\pi}\pi'^{2}}{\frac{27\pi'^{2}}{16m_{6}^{2}} - \Omega_{,\pi}} \ ,
\label{pipp}
\end{eqnarray}
where primes denote derivatives with respect to $y$. The junction conditions (\ref{covjc1}) and (\ref{covjc2}) can similarly be used to obtain the brane equations of motion. The $\pi$ junction condition (\ref{covjc2}) is trivial for a flat bulk and the $(\mu,\nu)$ components of (\ref{covjc1}) reduce to,
\begin{eqnarray}
	-\Omega_{,\pi_{0}}\pi'_{0} + \frac{9\pi'^{3}_{0}}{16m_{6}^{2}} = \frac{\Lambda}{2M_{5}^{3}} \ ,
\label{jc}
\end{eqnarray}
where the subscript 0 indicates that the function is evaluated at the brane position $y = 0$. We have further assumed that the matter energy-momentum tensor on the brane is a pure cosmological constant $\Lambda$, which we allow to be of any size, performing no fine-tuning like that usually required for the cosmological constant. In fact we would like $\Lambda$ to be large (TeV scale), since we know from particle physics experiments that such energy densities exist on our $4D$ brane. Note that, although we neglect other matter for simplicity, its inclusion would not affect our overall conclusions.

As a check, note that our junction condition~(\ref{jc}) is consistent with the decoupling limit result $\pi'_{0} = \Lambda/3M_{5}^{3}$ obtained in~\cite{deRham:2007xp,deRham:2010rw}. Indeed, in this limit $\Omega_{,\pi_{0}} \approx -3/2$. Moreover, introducing the canonically normalized $\pi^c = M_5^{3/2}\pi$, we see that the $\pi'^{3}$ term drops out in the limit $M_{5} \rightarrow \infty$, $m_{6} \rightarrow 0$ keeping $\Lambda_{6} = (m_{6}^{4}M_{5}^{3})^{1/7}$ fixed. Hence our junction condition~(\ref{jc}) reduces to the decoupling result in this limit.

It is easily seen that the bulk equation~(\ref{pipp}) allows for a first integral of motion
\be
- \Omega_{,\pi}\pi' + \frac{9\pi'^{3}}{16m_{6}^{2}} = {\rm constant}\,.
\ee
Comparing against the junction condition~(\ref{jc}) immediately fixes the integration constant in terms of $\Lambda$, and we obtain
\begin{eqnarray}
	- \Omega_{,\pi}\pi' + \frac{9\pi'^{3}}{16m_{6}^{2}} = \frac{\Lambda}{2M_{5}^{3}} \ .
\label{jcfirstint}
\end{eqnarray}
Notice that for suitable $\Omega$, (\ref{jcfirstint}) appears to admit a solution $\pi(y)$ for \textit{arbitrarily large} $\Lambda$. 

For example, suppose that $\Lambda$ is large and positive, and we choose $\Omega$ such that $\Omega_{,\pi} \rightarrow 0$ at large $\pi$ so that the cubic interaction term dominates everywhere, then this leads to a linear solution $\pi(y)$ increasing monotonically with $y$:
\be
\pi(y) \simeq \left(\frac{8m_6^2\Lambda}{9M_5^3}\right)^{1/3}y\,.
\label{pilin}
\ee
Since $\pi$ is non-singular for any finite $y$, the solution is well-defined everywhere. Therefore a flat brane solution is allowed for any tension.
Of course, consistency of the effective theory requires that $\pi' \ll M_5$. Since $\pi$ is suppressed by the tiny scale $m_6$, this is a
weak requirement: 
\be
\frac{\pi'}{M_5} \simeq \left(\frac{8m_6^2\Lambda}{9M_5^6}\right)^{1/3} =  \left(\frac{8}{9}\frac{m_6^2}{m_5^2} \frac{\Lambda}{M_4^4}\right)^{1/3}\ll 1\,,
\label{pi'bound}
\ee
where in the last step we have used~(\ref{m5m6}). Even with $\Lambda \sim M_4^4$, this can be satisfied provided $m_6\ll m_5$. A linearly growing $\pi(y)$ is also desirable from the point of view of quantum corrections to the $\pi$ Lagrangian. It is well-known that such corrections are of the form $(\Box\pi)^n$, that is, they always involve two derivatives per field, and hence vanish on a linear background. 

Note that the above remarks depend crucially on the cascading mechanism. If we let $m_{6} \rightarrow \infty$, thereby effectively decoupling the sixth dimension and turning off the cubic $\pi$ terms in~(\ref{5dcovcomplete}), then~(\ref{jcfirstint}) reduces to $-\Omega' = \Lambda/2M_{5}^{3}$, with solution $\Omega = -(\Lambda/2M_{5}^{3})y + c$. For $\Lambda > 0$, as assumed above, the integration constant $c$ must be positive since $\Omega$ must always be positive (since it is the coefficient of $R_{5}$ in the action). Hence $\Omega$ inevitably vanishes at some finite value of $y$ in this case, indicating strong coupling. (In Einstein frame, this corresponds to a naked singularity.) The cascading mechanism, therefore, is crucial in obtaining a flat brane solution for positive tension.

To gain further insight, we can translate to the Einstein frame: $g_{MN}^{\rm E} = \Omega^{2/3}\eta_{MN}$. In this frame, the brane extrinsic curvature is non-zero and is determined by the Israel junction condition. Focusing on its trace for simplicity, and assuming $\Omega_0 = 1$ without loss of generality, we have
\be
	K^{\rm E} = \frac{4}{3}\left(\frac{9\pi'^{3}_0}{16m_{6}^{2}} - \frac{\Lambda}{2M_{5}^{3}} \right)\,.
\label{KE1}
\ee
In the absence of the $\pi'^3$ term (corresponding to $m_6\rightarrow \infty$), the junction condition would imply $K^{\rm E}/M_5 \sim \Lambda/M_5^4$.
In turn, requiring that the curvature remains sub-Planckian, $K^{\rm E}\ll M_5$, would in turn impose a bound on the tension: $\Lambda < M_5^4$~\cite{Dvali:2002pe}. (Phenomenologically, $M_5$ must be less than $\sim {\rm MeV}$, so this bound would be rather stringent.) Instead, using~(\ref{jc}) and~(\ref{pi'bound}), we obtain
\be
	\frac{K^{\rm E}}{M_5} \approx  \Omega_{,\pi_{0}} \left(\frac{8m_6^2\Lambda}{9m_5^2M_4^4}\right)^{1/3}\,.
\ee
Again assuming $m_6\ll m_5$,  this allows a Planck-scale tension, $\Lambda \sim M_4^4$, while keeping $K^{\rm E}\ll M_5$. In other words, the $\pi'^3$ contribution in~(\ref{KE1}) neutralizes the dangerous $\Lambda$ term, leaving behind a much smaller curvature. This screening mechanism results in an effectively weak source for bulk gravity. This, however, also suggests that $\pi$ must be a source of negative energy to screen positive tension on the brane. This is not surprising since galileons are known to violate the usual energy conditions~\cite{Nicolis:2009qm}.

Thus at the background level our proposed screening mechanism displays many desirable features. To be physically viable, the action (\ref{5dcov}) must be perturbatively stable around a flat bulk solution. We study this issue in detail in the next section. Unfortunately, we will find that the theory propagates ghosts around the large-tension solution~(\ref{pilin}). More generally, the absence of ghost instabilities, combined with the requirement that the bulk solution is well-defined everywhere, places stringent constraints on the form of $\Omega$ and the allowed values of $\Lambda$ that can be degravitated. In Sec.~\ref{conclusions} we discuss possible ways to extend the framework to relax the stability constraints.


\section{Stability}
\label{ghosts}

In this section we study the stability of the degravitated solutions described above, by perturbing the complete Jordan frame action (\ref{5dcovcomplete}) to quadratic order around the flat bulk metric~(\ref{flatbulk}). To do so, it is convenient to work in the Arnowitt-Deser-Misner (ADM) coordinates \cite{Arnowitt:1962hi} with $y$ playing the role of a ``time'' variable,
\begin{eqnarray}
	{\rm d}s^{2}_{(5)} = N^{2}{\rm d}y^{2} + q_{\mu\nu} ({\rm d}x^{\mu} + N^{\mu}{\rm d}y) ({\rm d}x^{\nu} + N^{\nu}{\rm d}y) \ ,
\label{ADMmetric}
\end{eqnarray}
where $N$ denotes as usual the lapse function and $N_\mu$ the shift vector. Focusing on scalar perturbations, we use the gauge freedom to make $q_{\mu\nu}$ conformally flat
\be
	q_{\mu\nu} = e^{2\zeta(x^{\mu},y)} \eta_{\mu\nu} \ .
\ee
Moreover, we keep the brane at fixed position $y=0$. (This of course does not completely fix the gauge in the bulk, but is sufficient for our purposes.)
We perturb the lapse function, shift vector and scalar field respectively as
\bea
	N &=& 1 + \delta N \ , \\
	N_{\mu} &=& \partial_{\mu}\beta \ , \\
	\pi & = & \bar{\pi}(y) + \hat{\pi}(x^\mu,y) \ .
\eea
Similarly, all functions of $\pi$ (such as $\Omega(\pi)$) evaluated on the background will be denoted by a bar. (In particular, the background equations in Sec. \ref{flatbrane} only hold for the barred quantities $\bar{\pi}(y)$ and $\bar{\Omega}(y)$.)

After some integration by parts, carefully keeping track of boundary terms, the complete action at quadratic order is given by
\bea
	S_{\rm pert} & = & \frac{M_{5}^{3}}{2} \int_{\rm bulk} {\rm d}^{5}x \ \bar{\Omega} \Bigg[ 6(\partial\zeta)^{2} - 6 \left( \delta N + \frac{\bar{\Omega}_{,\pi}}{\bar{\Omega}}\hat{\pi} \right) (\partial^{2}\zeta) + 12\zeta'^{2} + 8\frac{\bar{\Omega}_{,\pi}}{\bar{\Omega}}\hat{\pi}'\zeta' + 8\frac{\bar{\Omega}_{,\pi\pi}}{\bar{\Omega}}\bar{\pi}'\hat{\pi}\zeta' \nonumber \\
	& & \ \ \ \ \ - \ 8\frac{\bar{\Omega}_{,\pi}}{\bar{\Omega}}\bar{\pi}'\delta N\zeta' - 2\frac{\bar{\Omega}_{,\pi}}{\bar{\Omega}}\delta N \partial^{2}\hat{\pi} + \frac{2}{\bar{\Omega}} \partial^{2}\beta (\bar{\Omega}_{,\pi}\bar{\pi}'\delta N - 3\bar{\Omega}\zeta' - \bar{\Omega}_{,\pi}\hat{\pi}' - \bar{\Omega}_{,\pi\pi}\bar{\pi}'\hat{\pi}) \Bigg] \nonumber \\
	& & - \ \frac{27M_{5}^{3}}{32m_{6}^{2}} \int_{\rm bulk} {\rm d}^{5}x \Big[ 2\bar{\pi}''(\partial\hat{\pi})^{2} - 2\bar{\pi}'^{2} \delta N \partial^{2} \hat{\pi} + 8\bar{\pi}'^{2} \left( \hat{\pi}'\zeta' - \bar{\pi}'\delta N\zeta' \right) \nonumber \\
	& & \ \ \ \ \ + \ 2\bar{\pi}'^{2} \partial^{2}\beta (\bar{\pi}'\delta N - \hat{\pi}') \Big] \nonumber \\
	& & + \ \frac{M_{4}^{2}}{4} \int_{\rm brane} {\rm d}^{4}x [ 6(\partial\zeta)^{2} ] - \frac{27M_{5}^{3}}{32m_{6}^{2}} \int_{\rm brane} {\rm d}^{4}x [2\bar{\pi}'(\partial\hat{\pi})^{2}]\,.
\label{Spert1}
\eea
Varying with respect to $\beta$ and $N$ yields the first-order momentum and Hamiltonian constraint equations, respectively,
\bea
	\delta N & = & \frac{\hat{\pi}'}{\bar{\pi}'} - \frac{\bar{\pi}''}{\bar{\pi}'^{2}}\hat{\pi} - \frac{2\bar{\Omega}}{\bar{\pi}'Z}\zeta' \ ,
\label{deltaN} \\
	\partial^{2}\beta & = & -\frac{2\bar{\Omega}}{\bar{\pi}'Z} \partial^{2}\zeta + \frac{1}{\bar{\pi}'}\partial^{2}\hat{\pi} + 4\zeta' \ ,
\label{del2beta}
\eea
where we have defined
\bea
	Z \equiv - \frac{2}{3}\bar{\Omega}_{,\pi} + \frac{9\bar{\pi}'^{2}}{8m_{6}^{2}} \ .
\eea
Since $\delta N$ and $\beta$ are Lagrange multipliers, either of the relations~(\ref{deltaN}) and~(\ref{del2beta}) can be substituted back into~(\ref{Spert1}). The resulting quadratic action is
\bea
	S_{\rm pert} & = & \frac{M_{5}^{3}}{2} \int_{\rm bulk} {\rm d}^{5}x \left[ -12\bar{\Omega}\zeta'^{2} + \frac{12\bar{\Omega}^{2}}{Z^{2}} \left( \frac{Z^{2}}{2\bar{\Omega}} + \frac{Z\bar{\Omega}_{,\pi}}{\bar{\Omega}} - \frac{9\bar{\pi}''}{8m_{6}^{2}} \right) (\partial\zeta)^{2} \right] \nonumber \\
	& & + \ 3M_{5}^{3} \int_{\rm brane} {\rm d}^{4}x \left[ \frac{\bar{\Omega}}{\bar{\pi}'} \hat{\pi}\partial^{2}\zeta - \frac{Z}{4\bar{\pi}'} (\hat{\pi}\partial^{2}\hat{\pi}) + \frac{\bar{\Omega}^{2}}{\bar{\pi}'Z} (\partial\zeta)^{2} \right] \nonumber \\
	& & + \ \frac{M_{4}^{2}}{4} \int_{\rm brane} {\rm d}^{4}x [ 6(\partial\zeta)^{2} ] - \frac{27M_{5}^{3}}{32m_{6}^{2}} \int_{\rm brane} {\rm d}^{4}x [2\bar{\pi}'(\partial\hat{\pi})^{2}]\,.
\label{Spert2}
\eea

Note that the bulk action does not depend on $\hat{\pi}$, consistent with the fact that it is pure gauge from the bulk perspective. For consistency, its source at the brane position must vanish. That is, we must set the variation of the brane action with respect to $\hat{\pi}$ to zero, thus obtaining
\bea
	\hat{\pi} = \frac{\left(\frac{2\bar{\Omega}}{Z}\right)\zeta}{1-\frac{9}{4m_{6}^{2}}\frac{\bar{\pi}'^{2}}{Z}}\,.
\eea
Using this solution in (\ref{Spert2}) yields the complete $\zeta-$action,
\bea
	S_{\zeta} & = & \frac{M_{5}^{3}}{2} \int_{\rm bulk} {\rm d}^{5}x \left[ -12\bar{\Omega}\zeta'^{2} + \frac{12\bar{\Omega}^{2}}{Z^{2}} \left( \frac{Z^{2}}{2\bar{\Omega}} + \frac{Z\bar{\Omega}_{,\pi}}{\bar{\Omega}} - \frac{9\bar{\pi}''}{8m_{6}^{2}} \right) (\partial\zeta)^{2} \right] \nonumber \\
	& & - \ 3M_{5}^{3} \int_{\rm brane} {\rm d}^{4}x \left( 1-\frac{9\bar{\pi}'^{2}}{4m_{6}^{2}Z} \right)^{-1} \frac{9\bar{\pi}'}{4m_{6}^{2}Z^{2}} \, (\partial\zeta)^{2} + \frac{M_{4}^{2}}{4} \int_{\rm brane} {\rm d}^{4}x [ 6(\partial\zeta)^{2} ]\,.
\label{zetaactionJF}
\eea
where we have set $\bar{\Omega} = 1$ on the brane, without loss of generality. As a check, we have repeated the bulk calculation in the Einstein frame, where the bulk geometry is warped, and obtained the same result. This calculation is presented in the Appendix.

In order for bulk perturbations to be ghost-free, the coefficient of $(\partial\zeta)^{2}$ must be negative:
\bea
	\frac{Z^{2}}{2\bar{\Omega}} + \frac{Z\bar{\Omega}_{,\pi}}{\bar{\Omega}} - \frac{9\bar{\pi}''}{8m_{6}^{2}} < 0 \ .
\label{noghostbulk}
\eea
This inequality involves $\bar{\Omega}$, $\bar{\pi}'$ and $\bar{\pi}''$. Using the background equations of motion~(\ref{pipp}) and~(\ref{jcfirstint}), we can eliminate $\bar{\pi}'$ and $\bar{\pi}''$ in terms of $\bar{\Omega}$ and its derivatives, as well as the brane tension $\Lambda$. Hence~(\ref{noghostbulk}) reduces to a second-order differential inequality for $\bar{\Omega}(\bar{\pi})$, which
constrains the allowed functions $\Omega(\pi)$ that can yield ghost-free solutions for a given value of $\Lambda$. More precisely, since~(\ref{jcfirstint}) is a cubic equation for $\bar{\pi}'$, we obtain up to three allowed differential inequalities for $\bar{\Omega}(\bar{\pi})$. The physically-allowed $\Omega(\pi)$ should not only satisfy the ghost-free inequality, but must also be positive-definite and well-defined for all $y > 0$ to avoid strong coupling.

We have studied this problem numerically. Since it is non-trivial to solve the differential inequality directly, we have instead tried various forms
for $\Omega(\pi)$ for different values of $\Lambda$, and checked whether these forms satisfied the ghost-free condition~(\ref{noghostbulk}) for each of the roots of~(\ref{jcfirstint}). For each root that satisfied~(\ref{noghostbulk}), we then solved~(\ref{jcfirstint}) for $\bar{\pi}(y)$, and hence checked whether $\bar{\Omega}(y)$ remained positive and well-defined everywhere. Some of the specific functional forms we have tried
include $\Omega = 1 \pm 3\pi/2, e^{\pm 3\pi/2}$ and $1- 3\pi/2 + 9\pi^2/8$. 

\begin{figure}[!h]
  \begin{center}
    \includegraphics[width=3.1in,angle=0]{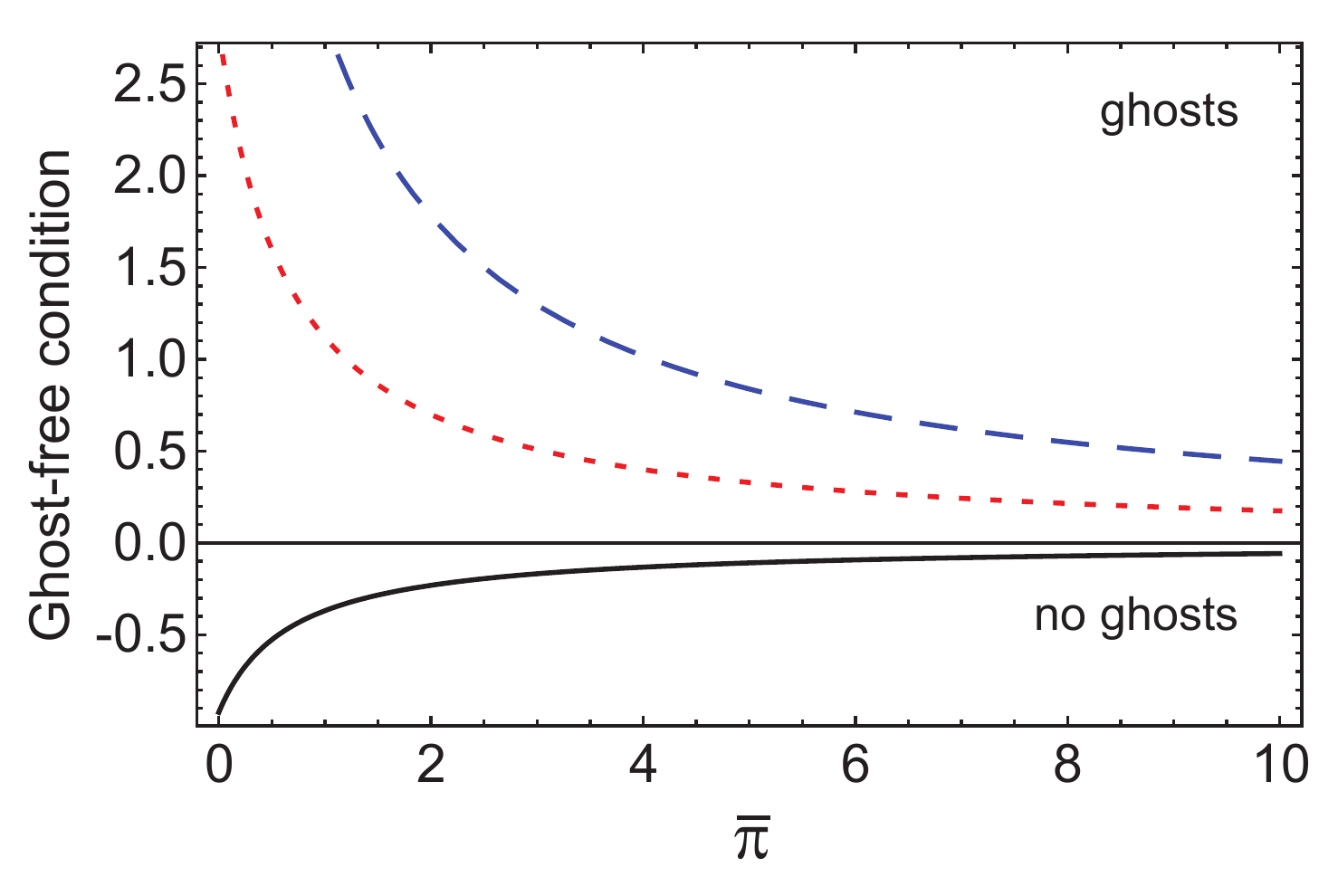}
    \includegraphics[width=3.03in,angle=0]{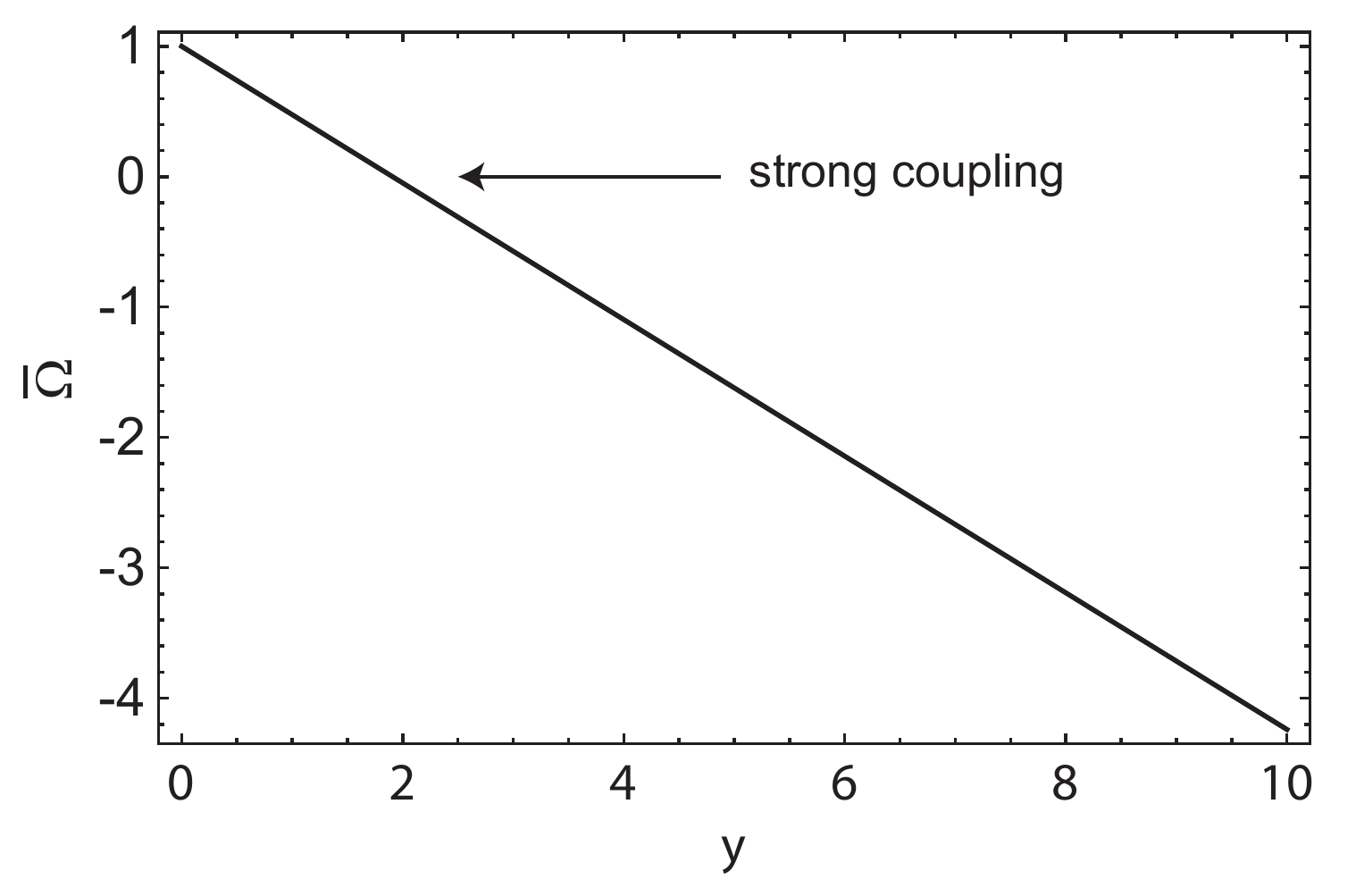}
    \caption{In the left panel, we plot the quantity $\frac{Z^{2}}{2\bar{\Omega}} + \frac{Z\bar{\Omega}_{,\pi}}{\bar{\Omega}} - \frac{9\bar{\pi}''}{8m_{6}^{2}}$ which appears in the ghost-free condition~(\ref{noghostbulk}) for $\Omega = 1 + 3\pi/2$ and $\Lambda = M_{6}^{4}$. The three curves correspond to the three roots of the cubic equation~(\ref{jcfirstint}) in $\bar{\pi}'/m_{6}$. The ghost-free condition requires $\frac{Z^{2}}{2\bar{\Omega}} + \frac{Z\bar{\Omega}_{,\pi}}{\bar{\Omega}} - \frac{9\bar{\pi}''}{8m_{6}^{2}} < 0$, hence only the black (solid) curve is free of ghost instabilities. In the right panel, we plot $\bar{\Omega}(y)$ for the ghost-free case. Since $\bar{\Omega}$ vanishes at finite $y$, corresponding to strong coupling, this solution is unphysical. We have found similar results for all positive values of $\Lambda$ and functional forms of $\Omega$ that we have tried.}
    \label{fig1}
  \end{center}
\end{figure}

For positive tension, $\Lambda > 0$, we were unable to find {\it any} $\Omega(\pi)$ that could simultaneously satisfy the ghost-free condition and remain everywhere well-defined and positive. For large tension, $\Lambda\gg M_6^4$, any real root of~(\ref{jcfirstint}) inevitably violates the ghost-free condition~(\ref{noghostbulk}). For small tension, $\Lambda\ll M_6^4$, it is possible to satisfy the ghost-free inequality, but the resulting $\Omega(y)$ either vanishes or becomes cuspy a finite distance from the brane. This is illustrated in Fig.~\ref{fig1} for the case $\Omega(\pi) = 1 + 3\pi/2$ and $\Lambda = M_6^4$.


\begin{figure}[!h]
  \begin{center}
    \includegraphics[width=3.1in,angle=0]{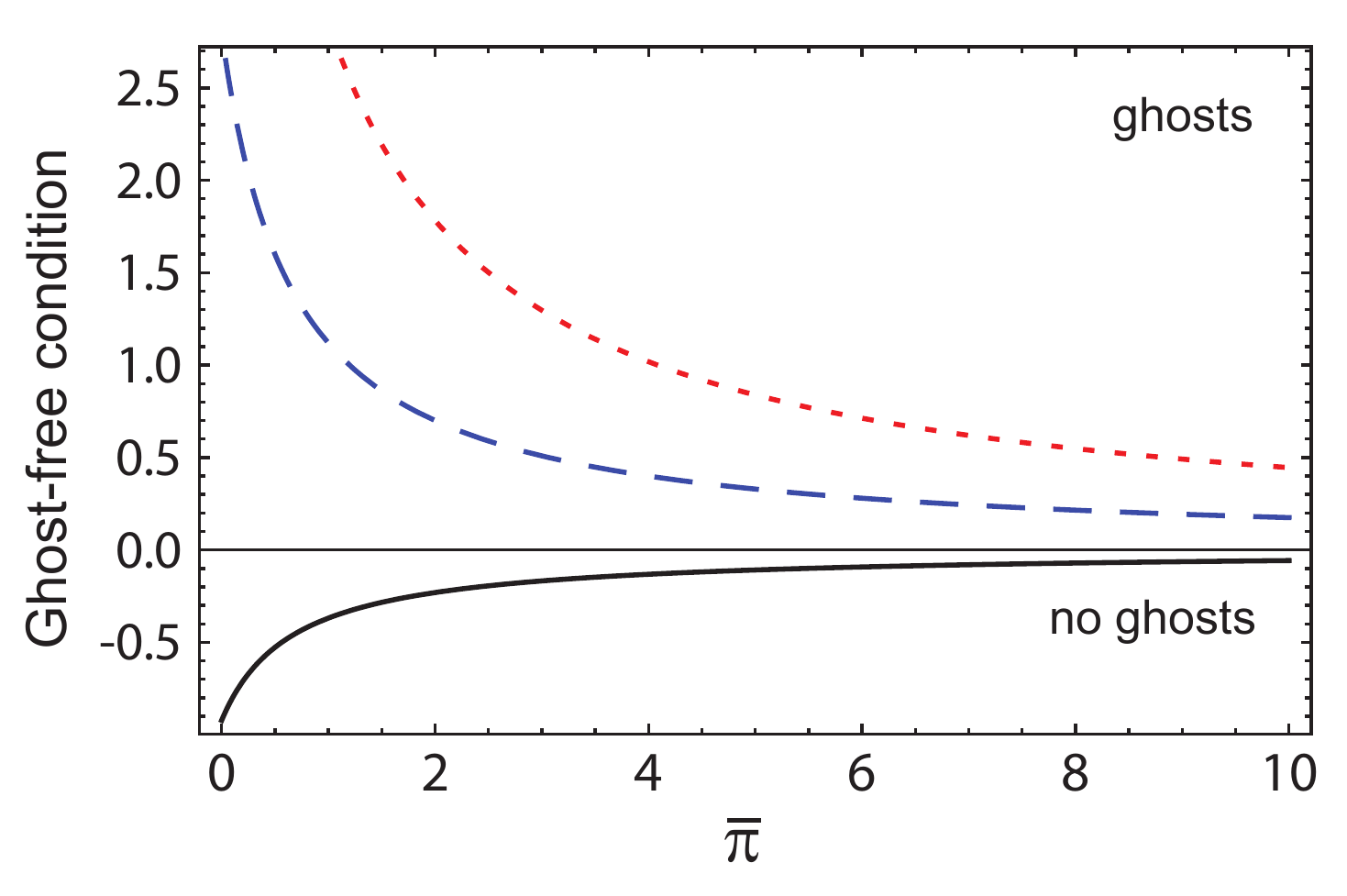}
    \includegraphics[width=2.95in,angle=0]{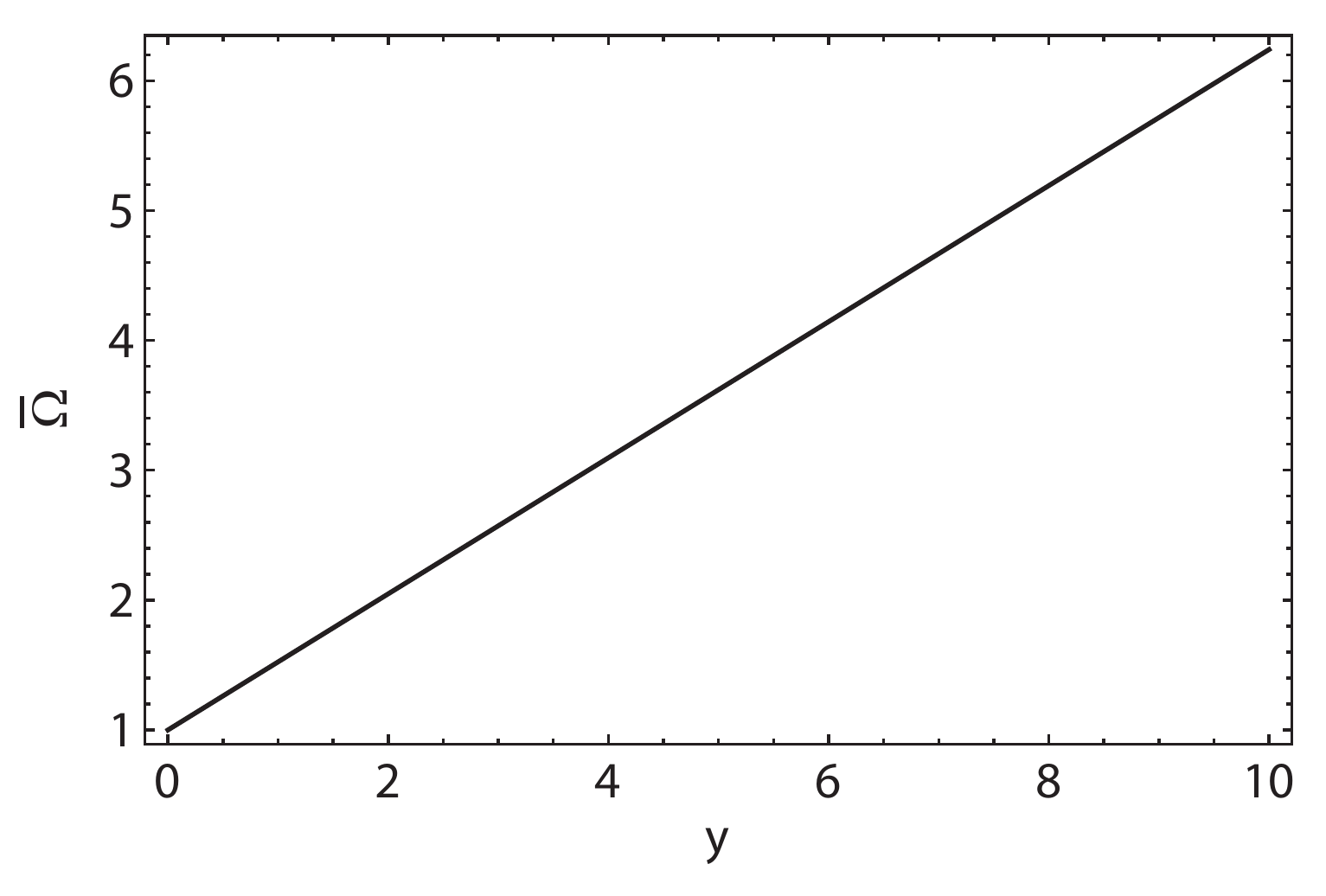}
    \caption{Same as Fig.~\ref{fig1}, except that $\Omega = 1 + 3\pi/2$ and $\Lambda = -M_{6}^{4}$. From the right panel, we see that $\bar{\Omega}(y)$ corresponding to the ghost-free branch is everywhere positive, hence this solution is physically viable.}
    \label{fig2}
  \end{center}
\end{figure}

For {\it negative} tension, $\Lambda < 0$, on the other hand, it is possible to find suitable $\Omega(\pi)$ that satisfy the ghost-free condition and are well-defined for all $y > 0$. Figure~\ref{fig2} illustrates this for $\Omega = 1 + 3\pi/2$ and $\Lambda = -M_6^4$. However, this is only the case for sufficiently small values of the tension, $|\Lambda|\lsim M_6^4$. For large values $|\Lambda| \gg M_6^4$, either the ghost-free condition cannot be satisfied or $\Omega(y)$ is ill-behaved. The existence of non-singular, ghost-free degravitated solutions, albeit with negative tension, is certainly a welcome feature of our $5D$ covariant framework. That said, these solutions do not connect to the parent $6D$ cascading framework, where the deficit angle solution requires a positive tension source.

Coming back to~(\ref{zetaactionJF}), there are other requirements that our degravitated solutions must satisfy. To avoid gradient instabilities in the extra dimension, the coefficient of $\zeta'^{2}$ must be negative, which is automatically true since $\Omega > 0$. Furthermore, from the ratio of the $\zeta'^{2}$ and $(\partial\zeta)^{2}$ terms we can infer the sound speed of propagation in the bulk:
\bea
	c_s^2 = \frac{-\frac{Z^{2}}{\bar{\Omega}}}{\frac{Z^{2}}{2\bar{\Omega}} + \frac{Z\bar{\Omega}_{,\pi}}{\bar{\Omega}} - \frac{9\bar{\pi}''}{8m_{6}^{2}}} \ ,
\eea
which is of course manifestly positive once~(\ref{noghostbulk}) is satisfied. Using this we can determine whether the propagation of perturbations is sub- or super-luminal. For the ghost-free example $\Omega = 1 + 3\pi/2$ and $\Lambda = - M_{6}^{4}$ shown in Fig.~\ref{fig2}, $c_s^2$ is sub-luminal everywhere.

Finally, the coefficient of $(\partial\zeta)^{2}$ on the brane must be negative, in order to avoid ghost instabilities: 
\bea
	Z_{0}^{2} - \frac{9m_{5}\bar{\pi}'_{0}}{2m_{6}^{2}} \left( 1-\frac{9\bar{\pi}'^{2}_{0}}{4m_{6}^{2}Z_{0}} \right)^{-1} < 0 \ .
\label{noghostbrane}
\eea
With $m_5 \geq m_6$, for instance, this condition is satisfied for the negative-tension example of Fig.~\ref{fig2}. As a check, we can compare this ghost-free condition with the stability bound~(\ref{ghostfreebound}) obtained both in the decoupling limit~\cite{deRham:2007xp} and in the full $6D$ cascading framework~\cite{deRham:2010rw}. In the decoupling limit with $\Omega = 1 - 3\pi/2$, where we expect agreement with the cascading results,~(\ref{noghostbrane}) indeed reduces to $\Lambda > 2m_{6}^{2}M_{4}^{2}/3$. 

Note that the absence of ghosts on the brane can always be achieved by adding a suitably-large kinetic term for $\pi$ on the brane, thereby 
modifying~(\ref{noghostbrane}) to a trivial condition. This intrinsic kinetic term would not affect the background solution nor the
bulk perturbation analysis. In this sense, the bulk ghost-free condition~(\ref{noghostbulk}) is a more robust constraint on the theory.


\section{Conclusions}
\label{conclusions}

Cascading gravity is an interesting approach to understanding dark energy as a manifestation of the presence of large extra dimensions. Unlike previous attempts, such as the DGP model, the propagators in cascading gravity are free of divergences, and the model has been found to be perturbatively ghost-free. Moreover, cascading gravity offers a promising arena for realizing degravitation: both in the codimension-2~~\cite{deRham:2007xp} and codimension-3~\cite{deRham:2009wb} cases, at least for small brane tension, the bulk geometry has been shown to be non-singular and asymptotically-flat, while the induced $4D$ geometry is flat. 

In this paper, we have studied a recently-proposed effective $5D$ action of cascading gravity in an attempt to obtain flat brane solutions. Our analysis has uncovered an intringuing screening mechanism that can shield bulk gravity from a large tension on the brane, resulting in a small brane extrinsic curvature. The brane remains flat for {\it arbitrarily large} tension, while the bulk is non-singular. Although this model offers an attractive mechanism to generalize extra-dimension dark energy models to higher codimensions without any fine-tuning, the stability analysis imposes stringent constraints. The bulk solution is perturbatively unstable for positive brane tension, while it is possible to find stable solutions for sufficiently small negative brane tension.

Our model agrees with earlier work in the weak-field limit, hence we do not contradict results that cascading gravity is indeed ghost-free. It does, however, raise the interesting question --- is there a fundamental difference between a theory with large extra dimensions and an effective $4D$ scalar-tensor theory of gravity? A complete answer to this question demands a more detailed analysis, which we leave to future work.

To improve stability, we are currently investigating the impact of including higher-order galileon terms for $\pi$ in the bulk, generalizing the results of~\cite{Nicolis:2008in} to $5D$. Preliminary results show that these higher-order terms still allow for flat brane solutions, while greatly alleviating the stability issues. In particular, ghost-free solutions are now possible with positive tension. However, demanding that gravity on the brane is approximately $4D$ on sufficiently large scales appears to impose an upper bound on the brane tension. The results of this ongoing analysis will be presented in detail elsewhere.


\section*{Acknowledgments}

NA would like to thank Joyce Byun for useful discussions. N.A., R.B., and M.T. are supported by NASA ATP Grant No. NNX08AH27G. N.A. and R.B.'s  research is also supported by NSF CAREER Grant No. AST0844825, NSF Grant No. PHY0555216, and by Research Corporation. The work of M.T. is also supported by NSF Grant No. PHY0930521 and by Department of Energy Grant No. DE-FG05-95ER40893-A020. M.T. is also supported by the Fay R. and Eugene L. Langberg chair. The work of J.K. is supported in part by the Alfred P. Sloan Foundation.


\appendix
\section*{Appendix: Alternative analysis of scalar perturbations}
\label{appendix}

\renewcommand{\theequation}{A\arabic{equation}}
\setcounter{equation}{0}

In this appendix we present an alternative derivation of the bulk $\zeta$-action in (\ref{zetaactionJF}), by performing the stability analysis in the Einstein frame: $g_{MN}^{\rm E} = \Omega^{2/3}g_{MN}^{\rm J}$. We define a warp factor $a_{E}(y) = \Omega^{1/3}(y)$ and a rescaled coordinate ${\rm d}y_{\rm E} = \Omega^{1/3} {\rm d}y$. Removing the subscripts ``$E$" for simplicity, the bulk metric in Einstein frame is 
\be
	{\rm d}s^{2}_{\rm bulk} = a^{2}(y) (-{\rm d}\tau^{2} + {\rm d}\vec{x}^{2}) + {\rm d}y^{2} \ .
\label{bulkmetricEF}
\ee
The Einstein frame bulk action is given by
\bea
	S_{\rm bulk} & = & \frac{M_{5}^{3}}{2} \int_{\rm bulk}{\rm d}^{5}x \sqrt{-g_{5}} \left[ R_{5} - \frac{4}{\Omega} \left( \frac{\Omega_{,\pi}^{2}}{\Omega} - \frac{2\Omega_{,\pi\pi}}{3} \right) (\partial\pi)^{2} + \frac{8\Omega_{,\pi}}{3\Omega} (\Box_{5}\pi) \right] \nonumber \\
	& & - \ \frac{27M_{5}^{3}}{32m_{6}^{2}} \int_{\rm bulk}{\rm d}^{5}x \sqrt{-g_{5}} (\partial\pi)^{2} \left[ \frac{1}{\Omega^{1/3}} \Box_{5}\pi - \frac{\Omega_{,\pi}}{\Omega^{4/3}} (\partial\pi)^{2} \right] \ .
\label{5dcovEF}
\eea
Varying with respect to the metric yields the Einstein equations, $M_{5}^{3}G_{MN} = T_{MN}^{\pi}$, where the $\pi$ stress-energy tensor, 
$T_{MN}^{\pi} = -(2/\sqrt{-g_{5}}) \delta S_{\pi}/\delta g^{MN}$, is given by,
\bea
	T_{MN}^{\pi} & = & \frac{2M_{5}^{3}}{3} \frac{\Omega_{,\pi}^{2}}{\Omega^{2}} \Big[ 2\partial_{M}\pi \partial_{N}\pi - g_{MN} (\partial\pi)^{2} \Big] + \ \frac{9M_{5}^{3}}{16m_{6}^{2}} \frac{\Omega_{,\pi}}{\Omega^{4/3}} \Big[ g_{MN} (\partial\pi)^{4} - 4\partial_{M}\pi \partial_{N}\pi (\partial\pi)^{2} \Big] \nonumber \\
\nonumber
	& & - \ \frac{27M_{5}^{3}}{16m_{6}^{2}} \Omega^{-1/3} \Bigg[ \partial_{(M}(\partial\pi)^{2} \partial_{N)}\pi - \frac{1}{2} g_{MN}\partial_{K}(\partial\pi)^{2}\partial^{K}\pi - \partial_{M}\pi \partial_{N}\pi \Box_{5}\pi \nonumber \\
	& & + \ \frac{\Omega_{,\pi}}{3\Omega} \partial_{M}\pi \partial_{N}\pi(\partial\pi)^{2} \Bigg] \ .
\eea
For the metric (\ref{bulkmetricEF}) with $\pi \equiv \pi(y)$, the $(5,5)$ and $(\mu,\nu)$ components of the Einstein equations give us the following background evolution equations,
\bea
	6H^{2} & = & \rho \ , \\
	3H' & = & -(\rho + p) \ ,
\eea
where
\bea
	\rho & = & \frac{2}{3} \left[ \left(\frac{\Omega_{,\pi}}{\Omega}\right)^{2} \pi'^{2} - \frac{27}{8m_{6}^{2}} \Omega^{-1/3} \left( \frac{\Omega_{,\pi}}{\Omega} \pi'^{4} - 3H\pi'^{3} \right) \right] \ , \\
	p & = & \frac{2}{3} \left[ \left(\frac{\Omega_{,\pi}}{\Omega}\right)^{2} \pi'^{2} - \frac{27}{32m_{6}^{2}} \Omega^{-1/3} \left( \frac{\Omega_{,\pi}}{\Omega} \pi'^{4} + 3\pi'^{2}\pi'' \right) \right] \ .
\eea
Here $H = a'/a$ is the $5D$ Hubble parameter, with $y$ playing the role of a ``time'' variable.

To study scalar perturbations, we use ADM coordinates (\ref{ADMmetric}) and choose comoving gauge: $q_{\mu\nu} = a^2(y)e^{2\zeta(x^\mu,y)}\eta_{\mu\nu}$ and $\pi = \pi(y)$. In this gauge we cannot assume that the brane is at fixed position, but this is of no consequence here as we focus solely on bulk perturbations. The action (\ref{5dcovEF}) can be rewritten using ADM variables as
\be
	S_{\rm bulk}  = S_{g} + S_{\pi}\,,
\label{Sbulk}
\ee
with
\bea
\nonumber
	S_{g} & = & \frac{M_{5}^{3}}{2} \int_{\rm bulk} {\rm d}^{5}x \sqrt{-q}\left[ NR_{4} + \frac{1}{N}\left(E^{2} - E_{\mu\nu}E^{\mu\nu} \right)\right] \ , \\
	S_{\pi} & = & \frac{M_{5}^{3}}{2} \int_{\rm bulk} {\rm d}^{5}x \sqrt{-q}N \left[ -\frac{4}{3}\left(\frac{\Omega_{,\pi}}{\Omega}\right)^{2} \frac{\pi'^{2}}{N^{2}} \right] \nonumber \\
	& -&  \frac{27M_{5}^{3}}{32m_{6}^{2}} \int_{\rm bulk} {\rm d}^{5}x \sqrt{-q}N \frac{\pi'^{2}}{N^{2}} \left[ \Omega^{-1/3} \left( \frac{2}{3}\frac{\pi'}{N}K - \frac{8}{9} \frac{\Omega_{,\pi}}{\Omega} \frac{\pi'^{2}}{N^{2}} \right) \right] \ ,
\label{Spi1}
\eea
where $E_{\mu\nu} = (q_{\mu\nu}' - D_\mu N_\nu - D_\nu N_\mu)/2 = NK_{\mu\nu}$. 

Expanded to second order in the perturbations, $\delta N = N - 1$ and $\delta {E^{\alpha}}_{\alpha} = {E^{\alpha}}_{\alpha} - 4H$, the scalar field action reduces to
\bea
	S_{\pi} & = & \frac{M_{5}^{3}}{2} \int_{\rm bulk} {\rm d}^{5}x \sqrt{-q}N \Bigg[ 3H'\frac{1}{N^{2}} + 3(4H^{2}+H') + \frac{1}{2}M^{4}(y) \delta N^{2} - \hat{M}^{3}(y) \delta {E^{\alpha}}_{\alpha} \delta N \Bigg] \ , \nonumber \\
\label{Spi2}
\eea
where
\bea
	M^{4}(y) & = & -\frac{27}{8m_{6}^{2}} \Omega^{-1/3} \left( -\frac{11}{3}\frac{\Omega_{,\pi}}{\Omega}\pi'^{4} + \pi'^{2}\pi'' + 12H\pi'^{3} \right) \ , \\
	\hat{M}^{3}(y) & = & -\frac{27}{8m_{6}^{2}} \Omega^{-1/3} \pi'^{3} \ .
\eea
Varying the complete bulk action with respect to $N^{\mu}$ and $N$ gives us the momentum and Hamiltonian constraint equations,
\bea
	D_{\alpha} \left[ \frac{2}{N}(E{\delta^{\alpha}}_{\beta} - {E^{\alpha}}_{\beta}) - \hat{M}^{3}\delta N{\delta^{\alpha}}_{\beta}  \right] = 0 \ ,
\label{constraint1} \\
	R_{4} - \frac{1}{N^{2}} (E^{2} - E_{\mu\nu}E^{\mu\nu}) - \frac{3}{N^{2}}H' + 3(4H^{2}+H') + M^{4}\delta N - \hat{M}^{3}\delta {E^{\alpha}}_{\alpha} = 0 \ .
\label{constraint2}
\eea
For scalar perturbations, $q_{\mu\nu} = a^2(y)e^{2\zeta(x^\mu,y)}\eta_{\mu\nu}$ and $N_{\mu} \equiv \partial_{\mu}\beta$, the first-order solutions to~(\ref{constraint1}) and (\ref{constraint2}) are given by
\bea
	\delta N & = & \frac{6\zeta'}{6H + \hat{M}^{3}} \ ,
\label{solution1} \\
	\Box_{4}\beta & = & \frac{6}{6H + \hat{M}^{3}} \frac{1}{a^{2}} \partial^{2}\zeta + \frac{-36H' + 48H\hat{M}^{3} + 4\hat{M}^{6} - 6M^{4}}{(6H + \hat{M}^{3})^{2}} \zeta' \ .
\label{solution2}
\eea
As usual, we only need to solve the constraint equations at first-order in the perturbations to obtain the quadratic Lagrangian for $\zeta$, since the second-order terms will multiply the unperturbed constraint equations, which vanish~\cite{Maldacena:2002vr}. Also note that here $\Box_{4}\beta = q^{\mu\nu}D_{\mu}D_{\nu}\beta$ whereas $\partial^{2}\zeta = \eta^{\mu\nu}\partial_{\mu}\partial_{\nu}\zeta$.

The quadratic action for $\zeta$ is obtained by plugging back the solutions (\ref{solution1}) and (\ref{solution2}) into the original action (\ref{Sbulk}), (\ref{Spi1}). We find that all of the $\Box_{4}\beta$ terms add up to a total derivative, hence the final Einstein frame $\zeta-$action is 
\bea
	S_{\zeta} & = & \frac{M_{5}^{3}}{2} \int_{\rm bulk} {\rm d}^{5}x \ a^{4} \left[ A(y) \zeta'^{2} + B(y) \frac{1}{a^{2}} (\partial\zeta)^{2} \right] \ ,
\label{zetaactionEF}
\eea
where
\bea
	A(y) & = & \frac{6(18H' - 24H\hat{M}^{3} - 2\hat{M}^{6} + 3M^{4})}{(6H + \hat{M}^{3})^{2}} \ , \\
	B(y) & = & \frac{6(18H' + 6H\hat{M}^{3} + \hat{M}^{6} + 3\partial_{y}\hat{M}^{3})}{(6H + \hat{M}^{3})^{2}} \ ,
\eea
and $(\partial\zeta)^{2} = \eta^{\mu\nu}\partial_{\mu}\zeta\partial_{\nu}\zeta$. 

We can transform the action (\ref{zetaactionEF}) back to the Jordan frame by using the transformations between Einstein frame variables (now denoted with a subscript ``$E$") and Jordan frame variables: $a_{\rm E} = \Omega^{1/3}$, ${\rm d}y_{\rm E} = \Omega^{1/3}{\rm d}y$, and $\zeta_{\rm E} = \zeta$. The result is
\bea
	S_{\zeta}^{\rm Jordan} & = & \frac{M_{5}^{3}}{2} \int_{\rm bulk} {\rm d}^{5}x \left[ -12\Omega\zeta'^{2} + \frac{12\Omega^{2}}{Z^{2}} \left( \frac{Z^{2}}{2\Omega} + \frac{Z\Omega_{,\pi}}{\Omega} - \frac{9\pi''}{8m_{6}^{2}} \right) (\partial\zeta)^{2} \right] \ ,
\eea
which matches with the bulk Jordan frame action in (\ref{zetaactionJF}).


\bibliographystyle{apsrev}
\bibliography{degravitation}

\begin{thebibliography}{46}
\expandafter\ifx\csname natexlab\endcsname\relax\def\natexlab#1{#1}\fi
\expandafter\ifx\csname bibnamefont\endcsname\relax
  \def\bibnamefont#1{#1}\fi
\expandafter\ifx\csname bibfnamefont\endcsname\relax
  \def\bibfnamefont#1{#1}\fi
\expandafter\ifx\csname citenamefont\endcsname\relax
  \def\citenamefont#1{#1}\fi
\expandafter\ifx\csname url\endcsname\relax
  \def\url#1{\texttt{#1}}\fi
\expandafter\ifx\csname urlprefix\endcsname\relax\def\urlprefix{URL }\fi
\providecommand{\bibinfo}[2]{#2}
\providecommand{\eprint}[2][]{\url{#2}}

\bibitem[{\citenamefont{Lukas et~al.}(1999)\citenamefont{Lukas, Ovrut, Stelle,
  and Waldram}}]{Lukas:1998yy}
\bibinfo{author}{\bibfnamefont{A.}~\bibnamefont{Lukas}},
  \bibinfo{author}{\bibfnamefont{B.~A.} \bibnamefont{Ovrut}},
  \bibinfo{author}{\bibfnamefont{K.~S.} \bibnamefont{Stelle}},
  \bibnamefont{and} \bibinfo{author}{\bibfnamefont{D.}~\bibnamefont{Waldram}},
  \bibinfo{journal}{Phys. Rev.} \textbf{\bibinfo{volume}{D59}},
  \bibinfo{pages}{086001} (\bibinfo{year}{1999}), \eprint{hep-th/9803235}.

\bibitem[{\citenamefont{Arkani-Hamed et~al.}(1998)\citenamefont{Arkani-Hamed,
  Dimopoulos, and Dvali}}]{ArkaniHamed:1998rs}
\bibinfo{author}{\bibfnamefont{N.}~\bibnamefont{Arkani-Hamed}},
  \bibinfo{author}{\bibfnamefont{S.}~\bibnamefont{Dimopoulos}},
  \bibnamefont{and} \bibinfo{author}{\bibfnamefont{G.~R.} \bibnamefont{Dvali}},
  \bibinfo{journal}{Phys. Lett.} \textbf{\bibinfo{volume}{B429}},
  \bibinfo{pages}{263} (\bibinfo{year}{1998}), \eprint{hep-ph/9803315}.

\bibitem[{\citenamefont{Antoniadis et~al.}(1998)\citenamefont{Antoniadis,
  Arkani-Hamed, Dimopoulos, and Dvali}}]{Antoniadis:1998ig}
\bibinfo{author}{\bibfnamefont{I.}~\bibnamefont{Antoniadis}},
  \bibinfo{author}{\bibfnamefont{N.}~\bibnamefont{Arkani-Hamed}},
  \bibinfo{author}{\bibfnamefont{S.}~\bibnamefont{Dimopoulos}},
  \bibnamefont{and} \bibinfo{author}{\bibfnamefont{G.~R.} \bibnamefont{Dvali}},
  \bibinfo{journal}{Phys. Lett.} \textbf{\bibinfo{volume}{B436}},
  \bibinfo{pages}{257} (\bibinfo{year}{1998}), \eprint{hep-ph/9804398}.

\bibitem[{\citenamefont{Randall and
  Sundrum}(1999{\natexlab{a}})}]{Randall:1999ee}
\bibinfo{author}{\bibfnamefont{L.}~\bibnamefont{Randall}} \bibnamefont{and}
  \bibinfo{author}{\bibfnamefont{R.}~\bibnamefont{Sundrum}},
  \bibinfo{journal}{Phys. Rev. Lett.} \textbf{\bibinfo{volume}{83}},
  \bibinfo{pages}{3370} (\bibinfo{year}{1999}{\natexlab{a}}),
  \eprint{hep-ph/9905221}.

\bibitem[{\citenamefont{Randall and
  Sundrum}(1999{\natexlab{b}})}]{Randall:1999vf}
\bibinfo{author}{\bibfnamefont{L.}~\bibnamefont{Randall}} \bibnamefont{and}
  \bibinfo{author}{\bibfnamefont{R.}~\bibnamefont{Sundrum}},
  \bibinfo{journal}{Phys. Rev. Lett.} \textbf{\bibinfo{volume}{83}},
  \bibinfo{pages}{4690} (\bibinfo{year}{1999}{\natexlab{b}}),
  \eprint{hep-th/9906064}.

\bibitem[{\citenamefont{Dvali et~al.}(2000)\citenamefont{Dvali, Gabadadze, and
  Porrati}}]{Dvali:2000hr}
\bibinfo{author}{\bibfnamefont{G.~R.} \bibnamefont{Dvali}},
  \bibinfo{author}{\bibfnamefont{G.}~\bibnamefont{Gabadadze}},
  \bibnamefont{and} \bibinfo{author}{\bibfnamefont{M.}~\bibnamefont{Porrati}},
  \bibinfo{journal}{Phys. Lett.} \textbf{\bibinfo{volume}{B485}},
  \bibinfo{pages}{208} (\bibinfo{year}{2000}), \eprint{hep-th/0005016}.

\bibitem[{\citenamefont{Rydbeck et~al.}(2007)\citenamefont{Rydbeck, Fairbairn,
  and Goobar}}]{Rydbeck:2007gy}
\bibinfo{author}{\bibfnamefont{S.}~\bibnamefont{Rydbeck}},
  \bibinfo{author}{\bibfnamefont{M.}~\bibnamefont{Fairbairn}},
  \bibnamefont{and} \bibinfo{author}{\bibfnamefont{A.}~\bibnamefont{Goobar}},
  \bibinfo{journal}{JCAP} \textbf{\bibinfo{volume}{0705}}, \bibinfo{pages}{003}
  (\bibinfo{year}{2007}), \eprint{astro-ph/0701495}.

\bibitem[{\citenamefont{Fang et~al.}(2008)}]{Fang:2008kc}
\bibinfo{author}{\bibfnamefont{W.}~\bibnamefont{Fang}} \bibnamefont{et~al.},
  \bibinfo{journal}{Phys. Rev.} \textbf{\bibinfo{volume}{D78}},
  \bibinfo{pages}{103509} (\bibinfo{year}{2008}), \eprint{0808.2208}.

\bibitem[{\citenamefont{Lombriser et~al.}(2009)\citenamefont{Lombriser, Hu,
  Fang, and Seljak}}]{Lombriser:2009xg}
\bibinfo{author}{\bibfnamefont{L.}~\bibnamefont{Lombriser}},
  \bibinfo{author}{\bibfnamefont{W.}~\bibnamefont{Hu}},
  \bibinfo{author}{\bibfnamefont{W.}~\bibnamefont{Fang}}, \bibnamefont{and}
  \bibinfo{author}{\bibfnamefont{U.}~\bibnamefont{Seljak}},
  \bibinfo{journal}{Phys. Rev.} \textbf{\bibinfo{volume}{D80}},
  \bibinfo{pages}{063536} (\bibinfo{year}{2009}), \eprint{0905.1112}.

\bibitem[{\citenamefont{Guo et~al.}(2006)\citenamefont{Guo, Zhu, Alcaniz, and
  Zhang}}]{Guo:2006ce}
\bibinfo{author}{\bibfnamefont{Z.-K.} \bibnamefont{Guo}},
  \bibinfo{author}{\bibfnamefont{Z.-H.} \bibnamefont{Zhu}},
  \bibinfo{author}{\bibfnamefont{J.~S.} \bibnamefont{Alcaniz}},
  \bibnamefont{and} \bibinfo{author}{\bibfnamefont{Y.-Z.} \bibnamefont{Zhang}},
  \bibinfo{journal}{Astrophys. J.} \textbf{\bibinfo{volume}{646}},
  \bibinfo{pages}{1} (\bibinfo{year}{2006}), \eprint{astro-ph/0603632}.

\bibitem[{\citenamefont{Nicolis and Rattazzi}(2004)}]{Nicolis:2004qq}
\bibinfo{author}{\bibfnamefont{A.}~\bibnamefont{Nicolis}} \bibnamefont{and}
  \bibinfo{author}{\bibfnamefont{R.}~\bibnamefont{Rattazzi}},
  \bibinfo{journal}{JHEP} \textbf{\bibinfo{volume}{06}}, \bibinfo{pages}{059}
  (\bibinfo{year}{2004}), \eprint{hep-th/0404159}.

\bibitem[{\citenamefont{Koyama}(2005)}]{Koyama:2005tx}
\bibinfo{author}{\bibfnamefont{K.}~\bibnamefont{Koyama}},
  \bibinfo{journal}{Phys. Rev.} \textbf{\bibinfo{volume}{D72}},
  \bibinfo{pages}{123511} (\bibinfo{year}{2005}), \eprint{hep-th/0503191}.

\bibitem[{\citenamefont{Gorbunov et~al.}(2006)\citenamefont{Gorbunov, Koyama,
  and Sibiryakov}}]{Gorbunov:2005zk}
\bibinfo{author}{\bibfnamefont{D.}~\bibnamefont{Gorbunov}},
  \bibinfo{author}{\bibfnamefont{K.}~\bibnamefont{Koyama}}, \bibnamefont{and}
  \bibinfo{author}{\bibfnamefont{S.}~\bibnamefont{Sibiryakov}},
  \bibinfo{journal}{Phys. Rev.} \textbf{\bibinfo{volume}{D73}},
  \bibinfo{pages}{044016} (\bibinfo{year}{2006}), \eprint{hep-th/0512097}.

\bibitem[{\citenamefont{Charmousis et~al.}(2006)\citenamefont{Charmousis,
  Gregory, Kaloper, and Padilla}}]{Charmousis:2006pn}
\bibinfo{author}{\bibfnamefont{C.}~\bibnamefont{Charmousis}},
  \bibinfo{author}{\bibfnamefont{R.}~\bibnamefont{Gregory}},
  \bibinfo{author}{\bibfnamefont{N.}~\bibnamefont{Kaloper}}, \bibnamefont{and}
  \bibinfo{author}{\bibfnamefont{A.}~\bibnamefont{Padilla}},
  \bibinfo{journal}{JHEP} \textbf{\bibinfo{volume}{10}}, \bibinfo{pages}{066}
  (\bibinfo{year}{2006}), \eprint{hep-th/0604086}.

\bibitem[{\citenamefont{Dvali et~al.}(2007{\natexlab{a}})\citenamefont{Dvali,
  Gabadadze, Pujolas, and Rahman}}]{Dvali:2006if}
\bibinfo{author}{\bibfnamefont{G.}~\bibnamefont{Dvali}},
  \bibinfo{author}{\bibfnamefont{G.}~\bibnamefont{Gabadadze}},
  \bibinfo{author}{\bibfnamefont{O.}~\bibnamefont{Pujolas}}, \bibnamefont{and}
  \bibinfo{author}{\bibfnamefont{R.}~\bibnamefont{Rahman}},
  \bibinfo{journal}{Phys. Rev.} \textbf{\bibinfo{volume}{D75}},
  \bibinfo{pages}{124013} (\bibinfo{year}{2007}{\natexlab{a}}),
  \eprint{hep-th/0612016}.

\bibitem[{\citenamefont{Gregory et~al.}(2007)\citenamefont{Gregory, Kaloper,
  Myers, and Padilla}}]{Gregory:2007xy}
\bibinfo{author}{\bibfnamefont{R.}~\bibnamefont{Gregory}},
  \bibinfo{author}{\bibfnamefont{N.}~\bibnamefont{Kaloper}},
  \bibinfo{author}{\bibfnamefont{R.~C.} \bibnamefont{Myers}}, \bibnamefont{and}
  \bibinfo{author}{\bibfnamefont{A.}~\bibnamefont{Padilla}},
  \bibinfo{journal}{JHEP} \textbf{\bibinfo{volume}{10}}, \bibinfo{pages}{069}
  (\bibinfo{year}{2007}), \eprint{0707.2666}.

\bibitem[{\citenamefont{Rubakov and Shaposhnikov}(1983)}]{Rubakov:1983bz}
\bibinfo{author}{\bibfnamefont{V.~A.} \bibnamefont{Rubakov}} \bibnamefont{and}
  \bibinfo{author}{\bibfnamefont{M.~E.} \bibnamefont{Shaposhnikov}},
  \bibinfo{journal}{Phys. Lett.} \textbf{\bibinfo{volume}{B125}},
  \bibinfo{pages}{139} (\bibinfo{year}{1983}).

\bibitem[{\citenamefont{Weinberg}(1989)}]{Weinberg:1988cp}
\bibinfo{author}{\bibfnamefont{S.}~\bibnamefont{Weinberg}},
  \bibinfo{journal}{Rev. Mod. Phys.} \textbf{\bibinfo{volume}{61}},
  \bibinfo{pages}{1} (\bibinfo{year}{1989}).

\bibitem[{\citenamefont{Gibbons et~al.}(2001)\citenamefont{Gibbons, Kallosh,
  and Linde}}]{Gibbons:2000tf}
\bibinfo{author}{\bibfnamefont{G.~W.} \bibnamefont{Gibbons}},
  \bibinfo{author}{\bibfnamefont{R.}~\bibnamefont{Kallosh}}, \bibnamefont{and}
  \bibinfo{author}{\bibfnamefont{A.~D.} \bibnamefont{Linde}},
  \bibinfo{journal}{JHEP} \textbf{\bibinfo{volume}{01}}, \bibinfo{pages}{022}
  (\bibinfo{year}{2001}), \eprint{hep-th/0011225}.

\bibitem[{\citenamefont{Dvali et~al.}(2003)\citenamefont{Dvali, Gabadadze, and
  Shifman}}]{Dvali:2002pe}
\bibinfo{author}{\bibfnamefont{G.}~\bibnamefont{Dvali}},
  \bibinfo{author}{\bibfnamefont{G.}~\bibnamefont{Gabadadze}},
  \bibnamefont{and} \bibinfo{author}{\bibfnamefont{M.}~\bibnamefont{Shifman}},
  \bibinfo{journal}{Phys. Rev.} \textbf{\bibinfo{volume}{D67}},
  \bibinfo{pages}{044020} (\bibinfo{year}{2003}), \eprint{hep-th/0202174}.

\bibitem[{\citenamefont{Arkani-Hamed et~al.}(2002)\citenamefont{Arkani-Hamed,
  Dimopoulos, Dvali, and Gabadadze}}]{ArkaniHamed:2002fu}
\bibinfo{author}{\bibfnamefont{N.}~\bibnamefont{Arkani-Hamed}},
  \bibinfo{author}{\bibfnamefont{S.}~\bibnamefont{Dimopoulos}},
  \bibinfo{author}{\bibfnamefont{G.}~\bibnamefont{Dvali}}, \bibnamefont{and}
  \bibinfo{author}{\bibfnamefont{G.}~\bibnamefont{Gabadadze}}
  (\bibinfo{year}{2002}), \eprint{hep-th/0209227}.

\bibitem[{\citenamefont{Dvali et~al.}(2007{\natexlab{b}})\citenamefont{Dvali,
  Hofmann, and Khoury}}]{Dvali:2007kt}
\bibinfo{author}{\bibfnamefont{G.}~\bibnamefont{Dvali}},
  \bibinfo{author}{\bibfnamefont{S.}~\bibnamefont{Hofmann}}, \bibnamefont{and}
  \bibinfo{author}{\bibfnamefont{J.}~\bibnamefont{Khoury}},
  \bibinfo{journal}{Phys. Rev.} \textbf{\bibinfo{volume}{D76}},
  \bibinfo{pages}{084006} (\bibinfo{year}{2007}{\natexlab{b}}),
  \eprint{hep-th/0703027}.

\bibitem[{\citenamefont{Dvali and Gabadadze}(2001)}]{Dvali:2000xg}
\bibinfo{author}{\bibfnamefont{G.~R.} \bibnamefont{Dvali}} \bibnamefont{and}
  \bibinfo{author}{\bibfnamefont{G.}~\bibnamefont{Gabadadze}},
  \bibinfo{journal}{Phys. Rev.} \textbf{\bibinfo{volume}{D63}},
  \bibinfo{pages}{065007} (\bibinfo{year}{2001}), \eprint{hep-th/0008054}.

\bibitem[{\citenamefont{Dubovsky and Rubakov}(2003)}]{Dubovsky:2002jm}
\bibinfo{author}{\bibfnamefont{S.~L.} \bibnamefont{Dubovsky}} \bibnamefont{and}
  \bibinfo{author}{\bibfnamefont{V.~A.} \bibnamefont{Rubakov}},
  \bibinfo{journal}{Phys. Rev.} \textbf{\bibinfo{volume}{D67}},
  \bibinfo{pages}{104014} (\bibinfo{year}{2003}), \eprint{hep-th/0212222}.

\bibitem[{\citenamefont{Gabadadze and Shifman}(2004)}]{Gabadadze:2003ck}
\bibinfo{author}{\bibfnamefont{G.}~\bibnamefont{Gabadadze}} \bibnamefont{and}
  \bibinfo{author}{\bibfnamefont{M.}~\bibnamefont{Shifman}},
  \bibinfo{journal}{Phys. Rev.} \textbf{\bibinfo{volume}{D69}},
  \bibinfo{pages}{124032} (\bibinfo{year}{2004}), \eprint{hep-th/0312289}.

\bibitem[{\citenamefont{de~Rham et~al.}(2008{\natexlab{a}})}]{deRham:2007xp}
\bibinfo{author}{\bibfnamefont{C.}~\bibnamefont{de~Rham}} \bibnamefont{et~al.},
  \bibinfo{journal}{Phys. Rev. Lett.} \textbf{\bibinfo{volume}{100}},
  \bibinfo{pages}{251603} (\bibinfo{year}{2008}{\natexlab{a}}),
  \eprint{0711.2072}.

\bibitem[{\citenamefont{de~Rham
  et~al.}(2008{\natexlab{b}})\citenamefont{de~Rham, Hofmann, Khoury, and
  Tolley}}]{deRham:2007rw}
\bibinfo{author}{\bibfnamefont{C.}~\bibnamefont{de~Rham}},
  \bibinfo{author}{\bibfnamefont{S.}~\bibnamefont{Hofmann}},
  \bibinfo{author}{\bibfnamefont{J.}~\bibnamefont{Khoury}}, \bibnamefont{and}
  \bibinfo{author}{\bibfnamefont{A.~J.} \bibnamefont{Tolley}},
  \bibinfo{journal}{JCAP} \textbf{\bibinfo{volume}{0802}}, \bibinfo{pages}{011}
  (\bibinfo{year}{2008}{\natexlab{b}}), \eprint{0712.2821}.

\bibitem[{\citenamefont{Corradini
  et~al.}(2008{\natexlab{a}})\citenamefont{Corradini, Koyama, and
  Tasinato}}]{Corradini:2007cz}
\bibinfo{author}{\bibfnamefont{O.}~\bibnamefont{Corradini}},
  \bibinfo{author}{\bibfnamefont{K.}~\bibnamefont{Koyama}}, \bibnamefont{and}
  \bibinfo{author}{\bibfnamefont{G.}~\bibnamefont{Tasinato}},
  \bibinfo{journal}{Phys. Rev.} \textbf{\bibinfo{volume}{D77}},
  \bibinfo{pages}{084006} (\bibinfo{year}{2008}{\natexlab{a}}),
  \eprint{0712.0385}.

\bibitem[{\citenamefont{Corradini
  et~al.}(2008{\natexlab{b}})\citenamefont{Corradini, Koyama, and
  Tasinato}}]{Corradini:2008tu}
\bibinfo{author}{\bibfnamefont{O.}~\bibnamefont{Corradini}},
  \bibinfo{author}{\bibfnamefont{K.}~\bibnamefont{Koyama}}, \bibnamefont{and}
  \bibinfo{author}{\bibfnamefont{G.}~\bibnamefont{Tasinato}},
  \bibinfo{journal}{Phys. Rev.} \textbf{\bibinfo{volume}{D78}},
  \bibinfo{pages}{124002} (\bibinfo{year}{2008}{\natexlab{b}}),
  \eprint{0803.1850}.

\bibitem[{\citenamefont{de~Rham et~al.}(2009)\citenamefont{de~Rham, Khoury, and
  Tolley}}]{deRham:2009wb}
\bibinfo{author}{\bibfnamefont{C.}~\bibnamefont{de~Rham}},
  \bibinfo{author}{\bibfnamefont{J.}~\bibnamefont{Khoury}}, \bibnamefont{and}
  \bibinfo{author}{\bibfnamefont{A.~J.} \bibnamefont{Tolley}},
  \bibinfo{journal}{Phys. Rev. Lett.} \textbf{\bibinfo{volume}{103}},
  \bibinfo{pages}{161601} (\bibinfo{year}{2009}), \eprint{0907.0473}.

\bibitem[{\citenamefont{de~Rham et~al.}(2010)\citenamefont{de~Rham, Khoury, and
  Tolley}}]{deRham:2010rw}
\bibinfo{author}{\bibfnamefont{C.}~\bibnamefont{de~Rham}},
  \bibinfo{author}{\bibfnamefont{J.}~\bibnamefont{Khoury}}, \bibnamefont{and}
  \bibinfo{author}{\bibfnamefont{A.~J.} \bibnamefont{Tolley}},
  \bibinfo{journal}{Phys. Rev.} \textbf{\bibinfo{volume}{D81}},
  \bibinfo{pages}{124027} (\bibinfo{year}{2010}), \eprint{1002.1075}.

\bibitem[{\citenamefont{Agarwal et~al.}(2010)\citenamefont{Agarwal, Bean,
  Khoury, and Trodden}}]{Agarwal:2009gy}
\bibinfo{author}{\bibfnamefont{N.}~\bibnamefont{Agarwal}},
  \bibinfo{author}{\bibfnamefont{R.}~\bibnamefont{Bean}},
  \bibinfo{author}{\bibfnamefont{J.}~\bibnamefont{Khoury}}, \bibnamefont{and}
  \bibinfo{author}{\bibfnamefont{M.}~\bibnamefont{Trodden}},
  \bibinfo{journal}{Phys. Rev.} \textbf{\bibinfo{volume}{D81}},
  \bibinfo{pages}{084020} (\bibinfo{year}{2010}), \eprint{0912.3798}.

\bibitem[{\citenamefont{Nicolis et~al.}(2009)\citenamefont{Nicolis, Rattazzi,
  and Trincherini}}]{Nicolis:2008in}
\bibinfo{author}{\bibfnamefont{A.}~\bibnamefont{Nicolis}},
  \bibinfo{author}{\bibfnamefont{R.}~\bibnamefont{Rattazzi}}, \bibnamefont{and}
  \bibinfo{author}{\bibfnamefont{E.}~\bibnamefont{Trincherini}},
  \bibinfo{journal}{Phys. Rev.} \textbf{\bibinfo{volume}{D79}},
  \bibinfo{pages}{064036} (\bibinfo{year}{2009}), \eprint{0811.2197}.

\bibitem[{\citenamefont{Luty et~al.}(2003)\citenamefont{Luty, Porrati, and
  Rattazzi}}]{Luty:2003vm}
\bibinfo{author}{\bibfnamefont{M.~A.} \bibnamefont{Luty}},
  \bibinfo{author}{\bibfnamefont{M.}~\bibnamefont{Porrati}}, \bibnamefont{and}
  \bibinfo{author}{\bibfnamefont{R.}~\bibnamefont{Rattazzi}},
  \bibinfo{journal}{JHEP} \textbf{\bibinfo{volume}{09}}, \bibinfo{pages}{029}
  (\bibinfo{year}{2003}), \eprint{hep-th/0303116}.

\bibitem[{\citenamefont{Chow and Khoury}(2009)}]{Chow:2009fm}
\bibinfo{author}{\bibfnamefont{N.}~\bibnamefont{Chow}} \bibnamefont{and}
  \bibinfo{author}{\bibfnamefont{J.}~\bibnamefont{Khoury}},
  \bibinfo{journal}{Phys. Rev.} \textbf{\bibinfo{volume}{D80}},
  \bibinfo{pages}{024037} (\bibinfo{year}{2009}), \eprint{0905.1325}.

\bibitem[{\citenamefont{Deffayet et~al.}(2009)\citenamefont{Deffayet,
  Esposito-Farese, and Vikman}}]{Deffayet:2009wt}
\bibinfo{author}{\bibfnamefont{C.}~\bibnamefont{Deffayet}},
  \bibinfo{author}{\bibfnamefont{G.}~\bibnamefont{Esposito-Farese}},
  \bibnamefont{and} \bibinfo{author}{\bibfnamefont{A.}~\bibnamefont{Vikman}},
  \bibinfo{journal}{Phys. Rev.} \textbf{\bibinfo{volume}{D79}},
  \bibinfo{pages}{084003} (\bibinfo{year}{2009}), \eprint{0901.1314}.

\bibitem[{\citenamefont{Silva and Koyama}(2009)}]{Silva:2009km}
\bibinfo{author}{\bibfnamefont{F.~P.} \bibnamefont{Silva}} \bibnamefont{and}
  \bibinfo{author}{\bibfnamefont{K.}~\bibnamefont{Koyama}},
  \bibinfo{journal}{Phys. Rev.} \textbf{\bibinfo{volume}{D80}},
  \bibinfo{pages}{121301} (\bibinfo{year}{2009}), \eprint{0909.4538}.

\bibitem[{\citenamefont{De~Felice and Tsujikawa}(2010)}]{DeFelice:2010pv}
\bibinfo{author}{\bibfnamefont{A.}~\bibnamefont{De~Felice}} \bibnamefont{and}
  \bibinfo{author}{\bibfnamefont{S.}~\bibnamefont{Tsujikawa}},
  \bibinfo{journal}{Phys. Rev. Lett.} \textbf{\bibinfo{volume}{105}},
  \bibinfo{pages}{111301} (\bibinfo{year}{2010}), \eprint{1007.2700}.

\bibitem[{\citenamefont{Mota et~al.}(2010)\citenamefont{Mota, Sandstad, and
  Zlosnik}}]{Mota:2010bs}
\bibinfo{author}{\bibfnamefont{D.~F.} \bibnamefont{Mota}},
  \bibinfo{author}{\bibfnamefont{M.}~\bibnamefont{Sandstad}}, \bibnamefont{and}
  \bibinfo{author}{\bibfnamefont{T.}~\bibnamefont{Zlosnik}},
  \bibinfo{journal}{JHEP} \textbf{\bibinfo{volume}{12}}, \bibinfo{pages}{051}
  (\bibinfo{year}{2010}), \eprint{1009.6151}.

\bibitem[{\citenamefont{Kachru et~al.}(2000)\citenamefont{Kachru, Schulz, and
  Silverstein}}]{Kachru:2000hf}
\bibinfo{author}{\bibfnamefont{S.}~\bibnamefont{Kachru}},
  \bibinfo{author}{\bibfnamefont{M.~B.} \bibnamefont{Schulz}},
  \bibnamefont{and}
  \bibinfo{author}{\bibfnamefont{E.}~\bibnamefont{Silverstein}},
  \bibinfo{journal}{Phys. Rev.} \textbf{\bibinfo{volume}{D62}},
  \bibinfo{pages}{045021} (\bibinfo{year}{2000}), \eprint{hep-th/0001206}.

\bibitem[{\citenamefont{York}(1972)}]{York:1972sj}
\bibinfo{author}{\bibfnamefont{J.}~\bibnamefont{York},
  \bibfnamefont{James~W.}}, \bibinfo{journal}{Phys. Rev. Lett.}
  \textbf{\bibinfo{volume}{28}}, \bibinfo{pages}{1082} (\bibinfo{year}{1972}).

\bibitem[{\citenamefont{Gibbons and Hawking}(1977)}]{Gibbons:1976ue}
\bibinfo{author}{\bibfnamefont{G.~W.} \bibnamefont{Gibbons}} \bibnamefont{and}
  \bibinfo{author}{\bibfnamefont{S.~W.} \bibnamefont{Hawking}},
  \bibinfo{journal}{Phys. Rev.} \textbf{\bibinfo{volume}{D15}},
  \bibinfo{pages}{2752} (\bibinfo{year}{1977}).

\bibitem[{\citenamefont{Dyer and Hinterbichler}(2009)}]{Dyer:2009yg}
\bibinfo{author}{\bibfnamefont{E.}~\bibnamefont{Dyer}} \bibnamefont{and}
  \bibinfo{author}{\bibfnamefont{K.}~\bibnamefont{Hinterbichler}},
  \bibinfo{journal}{JHEP} \textbf{\bibinfo{volume}{11}}, \bibinfo{pages}{059}
  (\bibinfo{year}{2009}), \eprint{0907.1691}.

\bibitem[{\citenamefont{Nicolis et~al.}(2010)\citenamefont{Nicolis, Rattazzi,
  and Trincherini}}]{Nicolis:2009qm}
\bibinfo{author}{\bibfnamefont{A.}~\bibnamefont{Nicolis}},
  \bibinfo{author}{\bibfnamefont{R.}~\bibnamefont{Rattazzi}}, \bibnamefont{and}
  \bibinfo{author}{\bibfnamefont{E.}~\bibnamefont{Trincherini}},
  \bibinfo{journal}{JHEP} \textbf{\bibinfo{volume}{05}}, \bibinfo{pages}{095}
  (\bibinfo{year}{2010}), \eprint{0912.4258}.

\bibitem[{\citenamefont{Arnowitt et~al.}(1962)\citenamefont{Arnowitt, Deser,
  and Misner}}]{Arnowitt:1962hi}
\bibinfo{author}{\bibfnamefont{R.~L.} \bibnamefont{Arnowitt}},
  \bibinfo{author}{\bibfnamefont{S.}~\bibnamefont{Deser}}, \bibnamefont{and}
  \bibinfo{author}{\bibfnamefont{C.~W.} \bibnamefont{Misner}}
  (\bibinfo{year}{1962}), \eprint{gr-qc/0405109}.

\bibitem[{\citenamefont{Maldacena}(2003)}]{Maldacena:2002vr}
\bibinfo{author}{\bibfnamefont{J.~M.} \bibnamefont{Maldacena}},
  \bibinfo{journal}{JHEP} \textbf{\bibinfo{volume}{05}}, \bibinfo{pages}{013}
  (\bibinfo{year}{2003}), \eprint{astro-ph/0210603}.

\end{thebibliography}

\end{document}